    \documentclass[preprint, aps, prb, onecolumn, superscriptaddress, floatfix, altaffilletter]{revtex4}
    \usepackage{graphicx}  
    \usepackage{dcolumn}   
    \usepackage{bm}        
    \usepackage{amssymb}  
    \usepackage{blindtext}
    \usepackage{pdfpages}
    \usepackage[nottoc]{tocbibind}
    \usepackage{bm}
    \usepackage{amsmath}
    \usepackage{multirow}
    \usepackage{hhline}
    \usepackage{physics}
    \usepackage{amsmath}
    \usepackage{tikz}
    \usetikzlibrary{shapes,shadows,arrows}
    \usepackage[caption=false,labelformat=parens]{subfig}
    \usepackage{wrapfig}
    \usepackage{caption}
    \usepackage{booktabs}
     \usepackage{float} 
    
    \usepackage{graphics}      

\begin{document}

\title{Energy Gap from Step Structure of the Analytically Inverted Non-Additive Kinetic Potential}
\author	   {Mojdeh Banafsheh}
\email          {mojdeh.banafsheh.co@gmail.com}
\affiliation    {Department of Physics, University of California, Merced, Merced, CA 95343, USA}
\author        {David A. Strubbe}
\email	{dstrubbe@ucmerced.edu}
\affiliation    {Department of Physics, University of California, Merced, Merced, CA 95343, USA}

\date{\today}

\begin{abstract}
The bandgap constitutes a challenging problem in density functional theory (DFT) methodologies. It is known that the energy gap values calculated by common DFT approaches are underestimated. The bandgap was also found to be related to the derivative discontinuity (DD) of the exchange-correlation potential in the Kohn-Sham formulation of DFT.
Several reports have shown that DD appears as a step on the potential curve. The step structure is a mandatory structure for aligning the KS energy levels in the ionization potentials in a dissociated molecule in both fragments and is a function of electron localisation. 
Reproducing the step in the DFT framework gives the charge transfer process and the correct energy gap and describes the source of dissociation. This step phenomenon has not yet been studied in the non-additive kinetic potential $v^{\text{NAD}}[\rho_A,\rho_B](\textbf{r})$, a key quantity used in embedding theories.  While $v^{\text{NAD}}[\rho_A,\rho_B](\textbf{r})$ is known to be difficult to approximate, in this work, we explain how an accurate energy gap can be produced from the analytically inverted $v^{\text{NAD}}[\rho_A,\rho_B](\textbf{r})$, even if we use the input densities calculated by the local and semi-local functionals.
 We used the precisely calculated $v^{\text{NAD}}[\rho_A,\rho_B](\textbf{r})$ reported in our previous publication [\textit{Phys. Rev. A} 106, 042812 (2022)] to produce the energy gap for some model systems and report in this work the promising accuracy of our results through the comparison with the results obtained from one of the most accurate calculations, OEP theory with the KLI local approximation. 
\end{abstract}

\maketitle
\section{Introduction}
In the context of Density Functional Theory (DFT), the energy gap can be calculated as the difference between the highest occupied molecular orbital (HOMO) energy and the lowest unoccupied molecular orbital (LUMO) energy. These energies are typically obtained from the eigenvalues of the Kohn-Sham equations, which are the fundamental equations of DFT. The energy obtained by such calculations does not produce an accurate result. To get the higher precision of the gap energy, based on the Kohn-Sham formulation of DFT, some orbital-dependent but more complicated calculations were produced that provided energy gap accurately. It would be extremely appreciated if, with some techniques, one could use 
the more straightforward methods, such as local or semilocal functional, to produce an accurate energy gap. 

The Kohn-Sham equations are solved self-consistently to obtain the electronic structure of a system. Once the electronic structure is obtained, the energies of the HOMO and LUMO orbitals can be determined. The energy gap, often referred to as the HOMO-LUMO gap, provides essential information about the electronic properties of the system, such as its electrical conductivity and optical properties.

It is important to note that the accuracy of the calculated energy gap in DFT depends on the choice of exchange-correlation functional, which is a key component of the theory. Different functionals can yield other predictions for the energy gap, and some functionals may perform better for specific types of systems or properties. In density functional theory (DFT), the accuracy of energy gap calculations can be influenced by various factors, including the choice of exchange-correlation function and the treatment of electronic excitations. While no single method can be considered universally ``most accurate" for all systems and properties, some approaches are known for providing reliable energy gap predictions in specific contexts. Here are a few methods commonly used for energy gap calculations in DFT:

Hybrid functionals, which incorporate a fraction of exact exchange and the standard DFT exchange-correlation functional, are often employed to improve the description of electronic band gaps. Examples of popular hybrid functionals include HSE06 (Heyd-Scuseria-Ernzerhof) and PBE0 (Perdew-Burke-Ernzerhof hybrid functional). These functionals can yield more accurate energy gaps for a wide range of systems, especially for semiconductors and insulators. 

The GW approximation, which combines Green's function theory with the screened Coulomb interaction, is a powerful method for calculating quasiparticle energies and excitation spectra. In the context of DFT, the GW method can be used to correct the DFT eigenvalues and improve the prediction of energy gaps, particularly for materials with strong electron correlation effects. However, GW calculations are computationally more demanding than standard DFT.
 
 Time-dependent DFT (TD-DFT) is a method that extends DFT to include the treatment of electronic excitations, such as optical transitions and electronic spectra. While TD-DFT can be less accurate for predicting fundamental gaps (HOMO-LUMO gaps) in some cases, it can provide reasonable estimates for excited-state energies and optical properties, making it helpful in studying the response of materials to light.
 
It is consequential to emphasise that the ``most accurate" method for energy gap calculations in DFT can depend on the specific properties of the system under study (e.g., band structure, electronic correlation effects) and the computational resources available. Researchers often need to consider trade-offs between accuracy and computational cost when selecting an appropriate method for their particular application. Additionally, empirical benchmarks and comparisons with experimental data are crucial for assessing the performance of different methods in practice.

While many successful approximations exist, they often
lack the derivative discontinuity (DD) \cite{perdew1982density,cohen2012challenges, baerends2013kohn,mori2014derivative, mosquera2014integer} of $E_{xc}[\rho](\textbf{r})$ concerning integer electron number, $N$. In some theories,
one of the properties of the accuracy of the theory is evaluated
when its $E_{xc}[\rho](\textbf{r})$ represents at DD that when added to the Kohn-Sham band gap, results in the fundamental band gap of the system. The fundamental gap  $E_g = I -A$ describes the difference between I, the ionisation potential (IP), and A, the electron affinity (EA). 

The presence of DD is expected in molecular dissociation 
when two atoms are stretched far apart\cite{makmal2011dissociation}. 
The atoms take on integer numbers of electrons to neutralise their charges,
so the total energy of the system, which is additive,
tends to show a DD relative to a number change
of electrons when one atom transfers its electron to another.

If any potential claims are exact, the DD has to appear
as a step on the curve of the potential solving
of a system of stretched atoms, open-shell systems, or
excited time-dependent systems. It is manifested in the
xc potential and appears in the form of a jump in the level of
the potential. The importance of the step structure (SS)
is in approaches that look for the tunnelling electrons and
charge transfer or electron excitation \cite{van1995step,helbig2009exact, gritsenko1996effect}. 

It is essential to consider the step structure in the approximations to ensure the accuracy of the methods. The step structure provides an accurate ground state electron density. 
It is a direct result of the necessary alignment of the KS energy levels (the ionisation
potentials) in the two fragments \cite{maitra2005undoing,levy1985constrained}.

In this article, after recalling in section \ref{OSS} the origin
of the step structure in stretched diatomic systems, the
physical interpretation of the step characteristic will be
illustrated in section \ref{StepHeight}. In this section, the link
between the height of the step and the derivative discontinuity
of $E_{xc}[\rho](\textbf{r})$ will be explained. The
same section will briefly describe the various types of step heights. Following
later, in section \ref{stepposition}, the relationship between the position
of the step occurring in the system and the associated ionisation energy will be discussed. Finally, the structure of the steps
appearing in non-additive potential is calculated precisely
of the analytical inversion of an admissible terrain pair
The charge density of the state is given in section \ref{stepVNAD}, and the number
of technical results for the model systems will be presented in section \ref{stepVNADResults}. 
In the same section, the precision of the energy gap is discussed,
and the comparison with other calculations is shown
in the tables relating to each model system.

\section{Origin of Step structure}\label{OSS}
A stretched heteronuclear diatomic molecule toward the molecular dissociation approaches into neutral atoms A and B. Atoms A and B in the form of isolated elements possessing different external potentials, so consequently different HOMO energies $\epsilon^A_{\rm HOMO}$ and $\epsilon^B_{\rm HOMO}$. In the exact Kohn-Sham formulation, HOMO energies must be equal to the first ionization energy:$\epsilon^A_{\rm HOMO} = -I_A$ and $\epsilon^B_{\rm HOMO} = -I_B$.
The variational principle for the total energy implies ensuring $\epsilon^A_{\rm HOMO} = \epsilon^B_{\rm HOMO}$ for when A and B are considered as constituent parts of a dissociated AB molecule.
Within the approximate theories of DFT, an artificial transfer of a fractional electron charge from
A to B is forced to ensure the equalisation of HOMO energies in the dissociated molecules. This leads to incorrect physics. Instead, in the exact formulations of DFT, this equality is obtained by the optimisation of $v_{xc}(\textbf{r})$ at the vicinity of one of the nuclei (Atom B for when the electronic charge has the potential of being transferred from A to B) that reads\cite{almbladh1985density}: 
\begin{eqnarray}\label{HomoAndIonisEqual}
\epsilon^A_{\rm HOMO} - \epsilon^B_{\rm HOMO} =I_B-I_A
\end{eqnarray}

The physical interpretation of this formula is that $v_{xc}(\textbf{r})$ represents a well for each atom. Through this equality, the $v_{xc}(\textbf{r})$ well of the atom with a higher electronegativity will be raised by $I_B-I_A$ accordingly to the  $v_{xc}(\textbf{r})$ well of the other atom.
This upshift is what is called  ``Step Structure'', characteristics to the exact $v_{xc}(\textbf{r})$. In some works of literature, it is also known as the ``counterionic field''\cite{gritsenko2006correct}. The name is deduced from the fact that this phenomenon will prevent
the electron density from flowing toward the more electronegative
atom. 
In DFT approximations, it is often found that atoms carry functional charges in a stretched molecule. The reason is that those approximations of the step structure are absent\cite{ruzsinszky2006spurious}. 
Step structure  of molecular $v_{xc}(\textbf{r})$ has been studied by different groups both analytically and numerically \cite{gritsenko1996molecular,gritsenko1996effect,gritsenko2006correct,karolewski2009polarizabilities,makmal2011dissociation, tempel2009revisiting,helbig2009exact, hellgren2012correlation, kraisler2015elimination,buijse1989j}.
Hodgson $et$ $al.$\cite{hodgson2016origin,hodgson2017interatomic} explained the origin of the step structure  from model systems for which the exact electron density is known. They showed that the steps form at points in the density when the probability of the existence of a charge in the local adequate ionisation energy of the electrons is nonzero. They also determined the shape of the step, its heights and position for ground state density both for time-dependent and time-independent systems. 
\subsection{Height of the step in a stretched diatomic molecule}\label{StepHeight}
The step height S for a system of two distant atoms is given by:
\begin{eqnarray}\label{StepHeight1}
S=I_R -I_L+\eta^{\rm HOMO}_R -\eta^{\rm HOMO}_L
\end{eqnarray}
where $I_L$ and $I_R$ are the ionisation energy of the left and right atoms, respectively and $\eta$ refers to the HOMO energy of molecular KS orbital localised around an atom. If the stretched molecule compromise two open-shell atoms, $\eta^{\rm HOMO}_R =\eta^{\rm HOMO}_L$ and Eq.[\ref{StepHeight1}] will simplify to $S=I_R -I_L$ which is the famous Almbladh and von Barth’s expression \cite{almbladh1985density}. If the overall ionisation potential  (IP) of the molecule is the IP of one of the atoms (for example, the one of the right atom), then $\eta^{\rm HOMO}_R =-I_R$ so the height of the step  becomes $S=\eta^{\rm HOMO}_L -I_L$. This latter can be zero or not, essentially. 

To interpret more clearly the functionality of the step, we may introduce the molecular energy $\eta^{\rm HOMO}_L$ in terms of the atomic energy $\epsilon^{\rm HOMO}_L$:
\begin{eqnarray}\label{MEnervsAEner}
\eta^{\rm HOMO}_L=\epsilon^{\rm HOMO}_L + S
\end{eqnarray}
Eq.[\ref{MEnervsAEner}] means that the molecular energy is elevated by the step height  relative to the atomic energy.
\subsection{Height of the step  in terms of the derivative discontinuity  of $E_{xc}[\rho](\textbf{r})$}
The Derivative Discontinuity  in the well-known approximations underestimate the gap \cite{perdew1992accurate, becke1988density, lee1988development,perdew1996generalized} underestimate the energy gap almost by $50\%$. The failure of those approximations may appear while calculating of charge transfer  in chemical processes. 
Derivative discontinuity  could be manifested as a uniform shift $\Delta$, in the level of exact Kohn-Sham potential in the position in real space when the ground state density integrates to a value that exceeds the integer value infinitesimally.
The shift $\Delta$ has a similar nature to the step height  explained previously: %
\begin{eqnarray}\label{StepDelta}
\Delta=\epsilon^{\rm HOMO}(N^{+}_0)-\epsilon^{\rm HOMO}(N^{-}_0)-(\epsilon^{\rm LUMO}-\epsilon^{\rm HOMO})
\end{eqnarray}
where $N_0$ is the expected integer value that the ground state density would integrate to in the absence of the derivative discontinuity  ($\int_V \rho_0(\textbf{r})d\textbf{r}$), and $\epsilon^{\rm HOMO}$ and $\epsilon^{\rm LUMO}$ are the highest occupied and lowest unoccupied Kohn-Sham eigenvalues, respectively. The charge $N_0$ varying infinitesimally by $\delta$ from integer is $N^{\pm}_0 = N_0 \pm \delta$.

As was already mentioned, the HOMO and LUMO Kohn-Sham eigenvalues are related to the ionisation potential  and  electron affinity\cite{yang2012derivative,levy1984exact,perdew1997comment,perdew1982density,harbola1999relationship}.
\begin{subequations}
\begin{alignat}{4}
 \epsilon^{\rm HOMO}(N^{+}_0) &= -A\\
 \epsilon^{\rm HOMO}(N^{-}_0) &= -I 
\end{alignat}
\label{subeqns}
\end{subequations}

It is important that $\Delta$ explains the correct energy gap in the form of Eq.[\ref{correctGap}].

\begin{eqnarray}\label{correctGap}
E_g=\Delta-(\epsilon^{\rm LUMO}-\epsilon^{\rm HOMO})
\end{eqnarray}
This is the case for the stretched molecules in which the atoms can not be considered as isolated. 
\subsection{ Step Height $S$ vs Step Height $\Delta$}
The two preceding phenomena, $\Delta$ and S, are generally
treated as independent and unrelated properties of the exact KS potential. Their differences are mainly related to the fact that $S$ is a quantity related to two fragments of the system in which the atoms could be approximated to the isolated elements, whereas $\Delta$ is the quantity that defines the correct energy gap of the entire system as a whole. 
 The EA and the LUMO
energy, yielding $\Delta$, are also absent from $S$. Convincingly, 
in a system, $S$ occurs at an integer of
electrons in a fixed location in the space, while the
$\Delta$ occurs when a small amount of charge is added to the whole system.
However, depending on the range of the system and the interaction between elements, the $S$ step could be related to the derivative discontinuity. 
The relation between the interatomic step, $S$, and the DD, $\Delta$ is published by Hodgson $et$ $al.$\cite{hodgson2017interatomic}.
\subsection{Position of the step } \label{stepposition}
Far from the localised electrons (around the midpoint of the interatomic axis) in subsystems, the density must decrease with the ionisation energy of the entire system. This is the case of a molecule of separate atoms with weakly overlapping atomic wavefunctions. 
Locally, a second charge in the effective ionisation energy exists away from the system if any subsystem density does not decrease with this energy. This local additional charge in the form of the step on $v_{xc}(\textbf{r})$  was initially observed by Perdew. \cite{perdew1985density} and then later by Makmal $et$ $al.$ \cite{makmal2011dissociation} in the exact
exchange potential for LiF, where they attribute the steps to
shifts in the Kohn-Sham eigenvalues. 

Thus, the step is expected to appear on the exact $v_{xc}(\textbf{r})$ to define the decay charge occurring in $\rho_0(\textbf{r})$ far from the nuclei. More about the position of the step structure can be found in Ref.\cite{tempel2009revisiting}.
	
			 \begin{eqnarray}\label{KS-Eq-One-Orb}
 				 v_{KS}(\textbf{r})=\frac{\nabla^2 \phi_i(\textbf{r})}{2\phi_i(\textbf{r})}+\epsilon_i
			 \end{eqnarray}

			Thus, the Kohn-Sham potential reads:

 			\begin{eqnarray}\label{KS-Eq-One-rho}
				v_{KS}(\textbf{r})= \frac{\nabla^2 \sqrt{\rho_i(\textbf{r})}}{2\sqrt{\rho_i(\textbf{r})}}+\epsilon_i 
  			\end{eqnarray}
For the model systems that can be solved by one-orbital formula (see EQ \ref{KS-Eq-One-Orb}), a spatial function in the form of $\tilde{I}(x)= \frac{1}{8\rho^2}(\frac{\partial \rho}{\partial x})^2$ could be defined that is sensitive to the ionisation energy for when the density decays asymptotically. As this asymptotic decay is manifested in the form of the step on the exact $v_{xc}(\textbf{r})$ so, $\tilde{I}(x)$ may predict the location in the space where the step occurs\cite{hodgson2016origin}. The detection of the step-through $\tilde{I}(x)$ is from the fact that in the vicinity of the nucleus $\tilde{I}(x)$ is equal to the corresponding ionisation energies whereas at the point where a change happens in the local ionisation energy (far from the nuclei), $\tilde{I}(x)$ shows a step. Both the height and the position of the step  were previously reported by the one-dimensional (1D) Heitler–London model wave function\cite{tempel2009revisiting}.
%
\subsection{Shape of the step }
The shift is spatially uniform only for $\delta \rightarrow 0^{+}$. In a finite
system, such as an atom or molecule, for any small but finite $\delta$,
$v_{KS}(\textbf{r})$ forms a “plateau” that elevates the level of the potential in
the vicinity of the nuclei. At the edge of the plateau, the level of
$v_{KS}(\textbf{r})$  must drop to $0$, forming a sharp spatial step. As $\delta$ vanishes,
the plateau extends over all space, becoming spatially uniform,
and its height approaches the value $\Delta$.

A plateau in the exact $v_{KS}(\textbf{r})$ is also observed for a different
physical scenario: a stretched diatomic molecule with an integer
number of electrons, $L\dots R$, which is one system consisting of
Atom $L$ and Atom $R$ with a large separation, $d$.

The plateau forms around one of the atoms, introducing an interatomic\cite{hellgren2012discontinuities,helbig2009exact,benitez2016kohn}, and ensuring the correct distribution of charge
throughout the system\cite{hofmann2012integer, perdew1990size,sagvolden2008discontinuity, levy1985constrained, makmal2011dissociation, fuks2011charge,gould2014delocalization, tempel2009revisiting,nafziger2015fragment,kohut2016origin,komsa2016elimination,li2015local}.In the general case, the
step height , $S$, is related to $I_R$ and $I_L$, the IPs of atoms $R$ and $L$,
respectively, as in Eq.[\ref{StepHeight1}].
%
\section{Step Structure  in $v^{\text{NAD}}[\rho_B,\rho_{tot}](\textbf{r})$}\label{stepVNAD}
Previously, Makmal $et$ $al.$\cite{makmal2011dissociation} published their work on molecular dissociation of a diatomic molecule with an all-electron Kohn-Sham solver DARSEC. They used the orbital formulation of OEP \cite{kummel2008orbital,sharp1953variational,talman1976optimized,grabo1997optimized,engel2003orbital} locally modified by KLI\cite{krieger1992construction}. Their main goal was to examine exact-exchange (EXX) Kohn-Sham potential. 
The $v_{xc}(\textbf{r})$ of semi-local functionals in molecules decays exponentially along with the density. The asymptotic behaviour of the exact $v_{xc}(\textbf{r})$ instead, is $-\frac{1}{r}$ \cite{almbladh1985exact,della2002asymptotic}. The incorrect asymptotic behaviour of the semi-local  $v_{xc}(\textbf{r})$ is the consequence of a self-interaction error. The self-interaction problem of semi-local potentials can be compensated by Perdew-Zunger self-interaction correction \cite{perdew1981self}. Even with such correction and calculations within the methods that are exact for all-electron calculations, the correct asymptotic behaviour of  $v_{xc}(\textbf{r})$ is not guaranteed\cite{vydrov2006scaling}.
It was shown by Makmal $et$ $al.$ that Kohn-Sham EXX plays an essential role in curing the problem of fractional molecular dissociation. They used stretched LiF molecules to study the performance of the EXX calculation. They achieved correct binding energy in their calculations while using the lowest-energy electronic configuration for different interatomic distances.
The correct binding energy was reflected by a plateau-like local Kohn-Sham $v_{xc}(\textbf{r})$ accompanied by two-step structures.

The steps that appeared locally on $v_{xc}(\textbf{r})$ bring the fact that the exact $v_{xc}(\textbf{r})$ of a diatomic molecule is not simply given by the sum of corresponding atomic potentials. This latter is always faithful even for considerable arbitrary interatomic distances. The reason is that one of the nuclear potentials is shifted by a constant due to the HOMO localised energy while the other atom lacks this shift. 

The density-dependent theories within DFT suffer from the lack of this precision in their $v_{xc}(\textbf{r})$ even the successful accurate approximations. Instead, the orbital-dependent approaches such as KLI can reflect the molecular dissociation through the derivative discontinuity  of the corresponding $E_{xc}(\textbf{r})$ at the level of the nonzero atomic shift of the potential. 

In our previous work \cite{banafsheh2022nuclear}, we showed that when locally
the integration of the ground state charge density tends
be $\leq 2$, the potential form of the analytical inversion approach is exact. We also discussed the potential walking wall
occurs in the position of the nuclei in the case where
the overlap charge density of a diatomic model tends to be zero at the midpoint of the atoms (Fig 3. b in \cite{banafsheh2022nuclear}); has
to the point that no steps occur, but instead, a small spike occurs.
pens where charge density integration crosses
slightly from zero.

Instead, when the charge densities overlap more significantly between atoms, we clearly observe the
occurrence of walking structures, as reported in Figure 5. b and 7. b in \cite{banafsheh2022nuclear}.

Knowing that the step structure is one of the main properties of an exact potential, We expect that the exact non-additive potential bi-functional shows the step structure spatially where the individual atomic potential shifts oppose. Hypothetically, such a step structure would produce an accurate energy gap in consequence.

None of the previously non-additive potential bi-functional was exact enough \cite{Banafsheh} to represent the step structure for relevant calculations. Instead, entirely or partially orbital-dependent calculations have the ability to illustrate the step structure.

Within DARSEC \cite{makmal2009fully}, we solved the same model systems with EXX locally approximated by KLI and also calculated $v^{\text{NAD}}[\rho_B,\rho_{tot}](\textbf{r})$ with a given ground state density.
We examined whether the step structure  appears spatially at the same position for both calculations. We also verified the relation between the height of the step  and the energy gap.
\section{Numerical Calculation}\label{NCSTEP}
The calculations are obtained on accurate numerical grids using the all-electron program package DARSEC \cite{makmal2009fully}. Consequently, we are restricted to computations of molecules with two
atomic centres with spherical symmetry. In DARSEC, the Kohn-Sham equations are solved self-consistently using the high-order finite difference approach \cite{fornberg1988generation,beck2000real}. This work set the stencil to 12 for the finite difference. 
A real-space grid based on prolate-spherical coordinates describes a system with two atomic centres. The grid is dense near two centres and increasingly sparse farther from the centres. Due to the cylindrical symmetry of diatomic molecules, the problem is reduced to a two-dimensional one. 
In the calculations for this work, the systems are defined within 15 Bohr of radius and the number of $115 \times 121$ grid points.  The $\rho_{tot}(\textbf{r})$ is the ground-state density of the systems performed with the LDA \cite{ceperley1980ground,perdew1992accurate}.

The calculations are done for two different sub-densities: $1)$ for when $\int \rho_B(\textbf{r})d\textbf{r}=2.0$, and for the case in which the charge density is not localised into an integer value at the vicinity of the nuclei, $\int \rho_B(\textbf{r})d\textbf{r}=1.5$.

   \begin{eqnarray}\label{Fermi-Dirac}
   F(z)= \frac{1}{e^{\alpha(z-z_0)}+1}	
   \end{eqnarray}
    By partitioning the whole density into two sub densities, we obtain: $\rho_B(\textbf{r})= F(z).\rho_{tot}(\textbf{r})$ and $\rho_A(\textbf{r})  =  \rho_{tot}(\textbf{r}) - \rho_B(\textbf{r})$. For the density localisation parameter $z_0$, see the details in section III.B of the paper Banafsheh et al. \cite{banafsheh2022nuclear}.

 For partitioning the ground state density numerically, we use a smooth distribution function  $0 \leq F(z)\leq 2$ that has no cusps and respects the smoothness of the function explained in Ref. \cite{banafsheh2022nuclear}. The choice for such a function used for the reported result is the Fermi-Dirac distribution function that changes smoothly from value one to zero (Eq.[\ref{Fermi-Dirac}]). This latter was realised within a binary-search algorithm to localise the density around one nucleus.
 
 All the calculations for the exact and the approximated theories were performed based on the same choice of the abovementioned parameters.

\section{Results and Discussion}\label{stepVNADResults}

\begin{figure}[h]
\begin{tikzpicture}
	\node [anchor=north west] (imgA) at (-0.275\linewidth,.90\linewidth)
			{\includegraphics[width=0.8\linewidth]{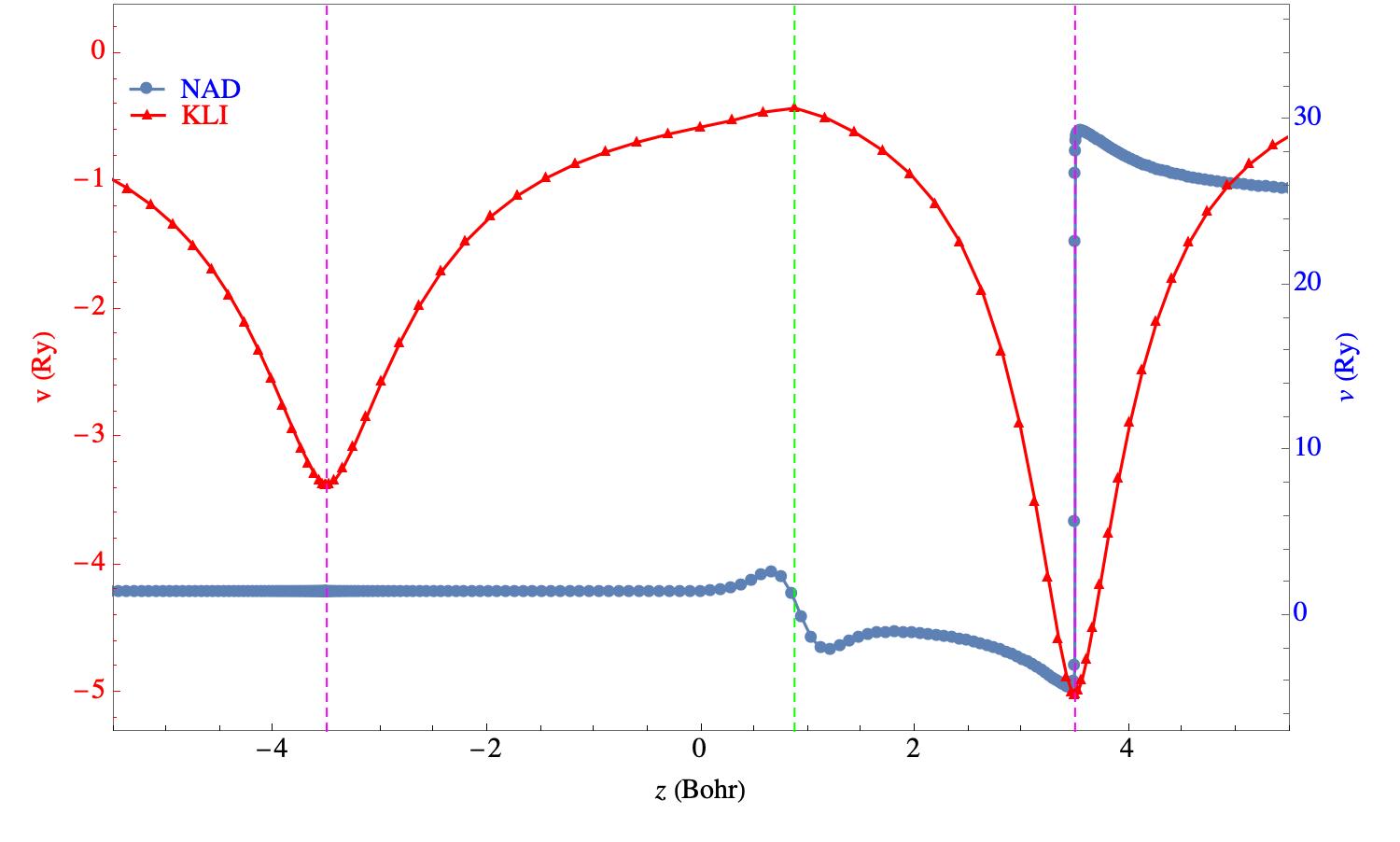}};
   \node [anchor=north west] (imgC) at (-0.275\linewidth,.35\linewidth)
            {\includegraphics[width=0.8\linewidth]{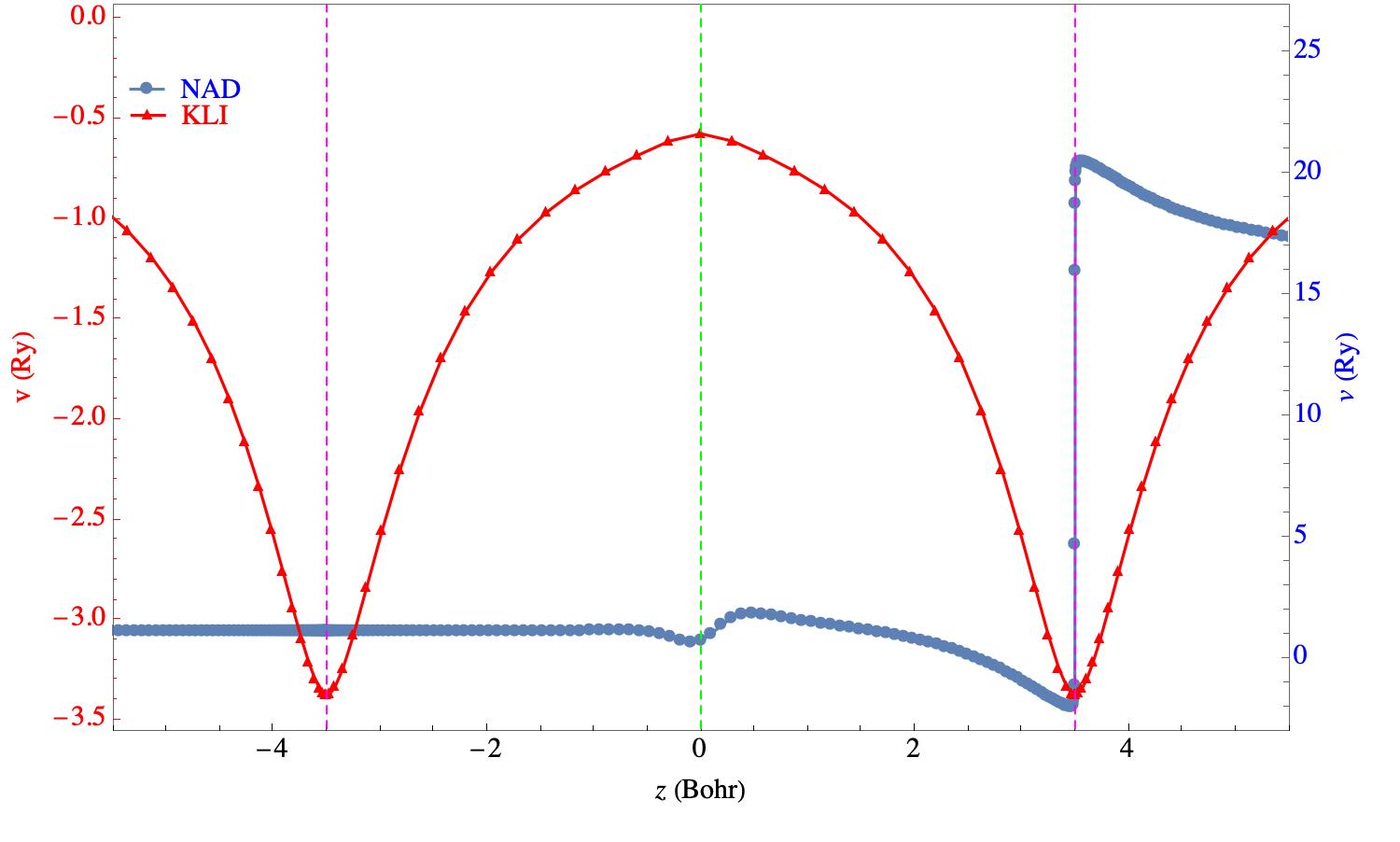}};
    \draw [anchor=north west] (-0.2\linewidth, .95\linewidth) node {(a) {\fontfamily{Arial}\selectfont {}}};
    \draw [anchor=north west] (-0.2\linewidth, .41\linewidth) node {(b) {\fontfamily{Arial}\selectfont {}}};
\end{tikzpicture}
\captionsetup{justification=raggedright,singlelinecheck=false}
\caption{  In search of the  Step Structure on the exact $v^{\text{NAD}}[\rho_B,\rho_{tot}(\textbf{r})]$ from LDA density input through the comparison of it with the location of the Step Structure on the curve of the $v^{\text{OEP-KLI}}_{xc}(\textbf{r})$; a) System: HeLi$^+$ where He is at $(0,0,-3.5)$ Bohr and Li is at $(0,0,3.5)$ Bohr; Blue line: $v^{\text{NAD}}[\rho_B,\rho_{tot}(\textbf{r})]$; Red Line:  $v^{\text{OEP-KLI}}_{xc}(\textbf{r})$; Vertical brown line: nuclei; Vertical green line: the location of the Step.
b) System: He-He where He is at $(0,0,-3.5)$ Bohr and He is at $(0,0,3.5)$ Bohr; Blue line: $v^{\text{NAD}}[\rho_B,\rho_{tot}(\textbf{r})]$; Red Line:  $v^{\text{OEP-KLI}}_{xc}(\textbf{r})$; Vertical brown line: nuclei; Vertical green line: the location of the Step.}
\label{StepKLIHeLi}
 \end{figure}

We seek for the SS in exact $v^{\text{NAD}}[\rho_B,\rho_{tot}](\textbf{r})$  analytically inverted from a ground state density and a partitioned sub-density integrating to two spin-compensated charge density for two diatomic model systems one heteronuclear and one homonuclear. The $v^{\text{NAD}}[\rho_B,\rho_{tot}](\textbf{r})$ appears in both cases with the SS in the space, where the overlap between two sub-densities is maximal. In Fig.[\ref{StepKLIHeLi}.a] and in Fig.[\ref{StepKLIHeLi}.b] we compared the related $v^{\text{NAD}}[\rho_B,\rho_{tot}](\textbf{r})$ with $v^{KLI}_{xc}(\textbf{r})$ for a heteronuclear and homonuclear model systems respectively. 

\begin{figure}[h]
\begin{tikzpicture}
	\node [anchor=north west] (imgA) at (-0.275\linewidth,.90\linewidth)
			{\includegraphics[width=0.8\linewidth]{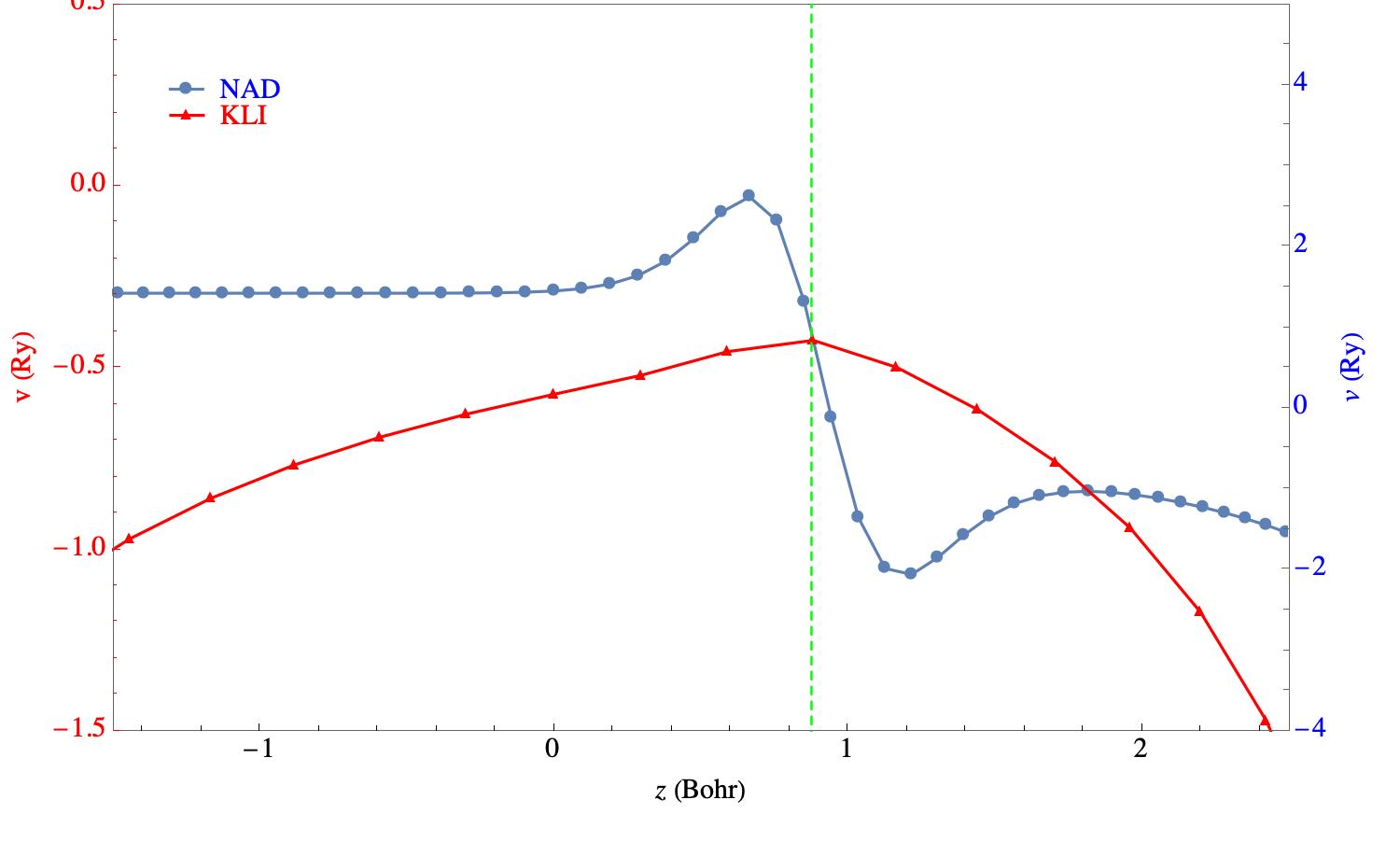}};
   \node [anchor=north west] (imgC) at (-0.275\linewidth,.35\linewidth)
            {\includegraphics[width=0.8\linewidth]{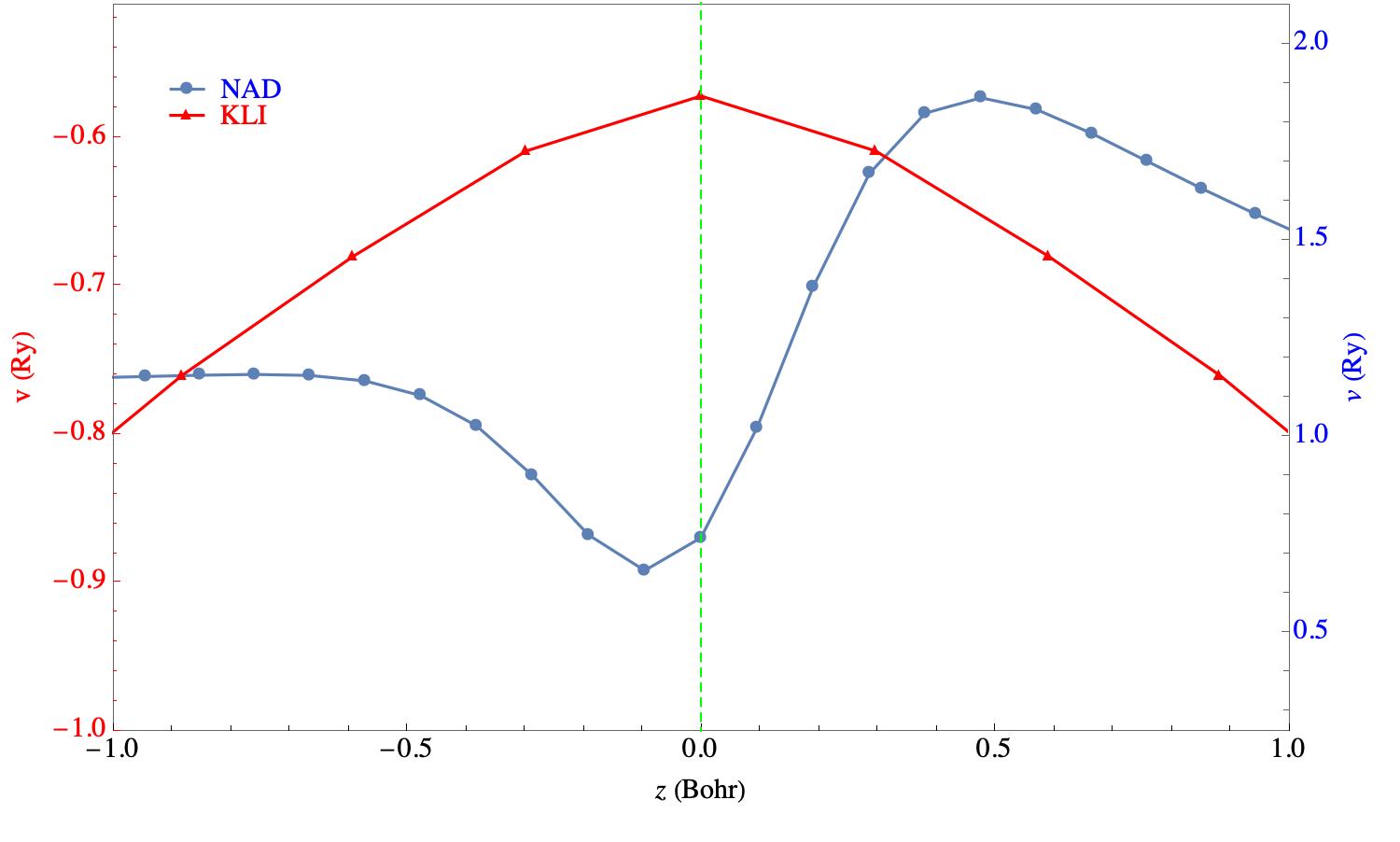}};
    \draw [anchor=north west] (-0.2\linewidth, .95\linewidth) node {(a) {\fontfamily{Arial}\selectfont {}}};
    \draw [anchor=north west] (-0.2\linewidth, .41\linewidth) node {(b) {\fontfamily{Arial}\selectfont {}}};
\end{tikzpicture}
\captionsetup{justification=raggedright,singlelinecheck=false}
\caption{  Zoom-in representation of the potentials around the overlap region between $\rho_A(\textbf{r})$ and $\rho_B(\textbf{r})$. a) System: HeLi$^+$ where He is at $(0,0,-3.5)$ Bohr and Li is at $(0,0,3.5)$ Bohr; Blue line: $v^{\text{NAD}}[\rho_B,\rho_{tot}(\textbf{r})]$; Red Line:  $v^{\text{OEP-KLI}}_{xc}(\textbf{r})$; Vertical green line: the location of the Step.
b) System:He-He where He is at $(0,0,-3.5)$ Bohr and Li is at $(0,0,3.5)$ Bohr; Blue line: $v^{\text{NAD}}[\rho_B,\rho_{tot}(\textbf{r})]$; Red Line:  $v^{\text{OEP-KLI}}_{xc}(\textbf{r})$; Vertical green line: the location of the Step.}
\label{ZoomedStepKLIHeLi}
 \end{figure}

The DD  in $v^{\text{NAD}}[\rho_B,\rho_{tot}](\textbf{r})$ appears spatially at the same position in which the SS appears on $v^{KLI}_{xc}(\textbf{r})$ in both models. In Figs. [\ref{ZoomedStepKLIHeLi}.a] and [\ref{ZoomedStepKLIHeLi}.b] the DD  of potentials are zoomed, and obviously, it doesn't appear easy to deduce the  height of the SS from the $v^{KLI}_{xc}(\textbf{r})$. Instead, the $v^{\text{NAD}}[\rho_B,\rho_{tot}](\textbf{r})$ shows more information about the behaviour of the related ground state density where the charge number varies infinitesimally from the integer where the sub-densities overlap. 

The zoomed-in plot on the overlap region in Fig.[\ref{ZoomedStepKLIHeLi}.a] and Fig.[\ref{ZoomedStepKLIHeLi}.b] shows that compared to $v^{KLI}_{xc}(\textbf{r})$, the step appears at the vicinity of the inflection point of the  in the  $v^{\text{NAD}}[\rho_B,\rho_{tot}](\textbf{r})$ closer to the first argument of the potential, here, $\rho_B(\textbf{r})$. This is true for both the heteronuclear and homonuclear model systems. This means the second derivative nor the third order of the $v^{\text{NAD}}[\rho_B,\rho_{tot}](\textbf{r})$ is zero where the step happens regarding the step position  from KLI. 

We need to dig more into the results of the heteronuclear system. Still, the step position  at first glance is expected to be exactly at the middle of the interatomic distance for a homonuclear model. 
In fact, it is not exactly the $v^{\text{NAD}}[\rho_B,\rho_{tot}](\textbf{r})$ that tells us about the exact position of the step. The difference between the exact analytically inverted potential (so-noted as $v^{\text{EXACT}}_s[\rho_{A/B}](\textbf{r})$) of the density of the separated systems ($v^{\text{EXACT}}_s[\rho_B](\textbf{r})$ or $v^{\text{EXACT}}_s[\rho_A](\textbf{r})$) can provide the exact position of the SS. In the end, what is obvious and the $v^{\text{NAD}}[\rho_B,\rho_{tot}](\textbf{r})$ or $v^{\text{NAD}}[\rho_A,\rho_{tot}](\textbf{r})$ have to reflect is that the inflection point of the curve occurs in the vicinity of the SS in the exact potential. In section \ref{GapFromVNAD}, the exact position of the SS from analytically inverted potential together with the correct energy gap will be discussed. 

\subsection{The Exact Position of the Step }
The origin of the exact position of the Step is based on the information carried by the SS and is predicted to happen in the curve of the $v^{\text{NAD}}[\rho_B,\rho_{tot}(\textbf{r})]$ where the overlap between $\rho_B(\textbf{r})$ and $\rho_A(\textbf{r})$ occurs to be maximal. The $\rho_A(\textbf{r})=\rho_{tot}(\textbf{r})-\rho_B(\textbf{r})$ where for the specific system models used in this work the $\int \rho_A(\textbf{r})d\textbf{r}=2$. 

Although the position of the step accurately appeared on the $v^{\text{NAD}}[\rho_B,\rho_{tot}(\textbf{r})]$, its exact position remains ambiguous on the curve. If the exact position of the step is directly related to the difference of the sub-densities, then the step has to happen on the local critical point of the curve $\Delta v(\textbf{r})]=v_s[\rho_A](\textbf{r})-v_s[\rho_B](\textbf{r})$.

\begin{figure}[h]
\begin{tikzpicture}
	\node [anchor=north west] (imgA) at (-0.275\linewidth,.90\linewidth)
			{\includegraphics[width=0.8\linewidth]{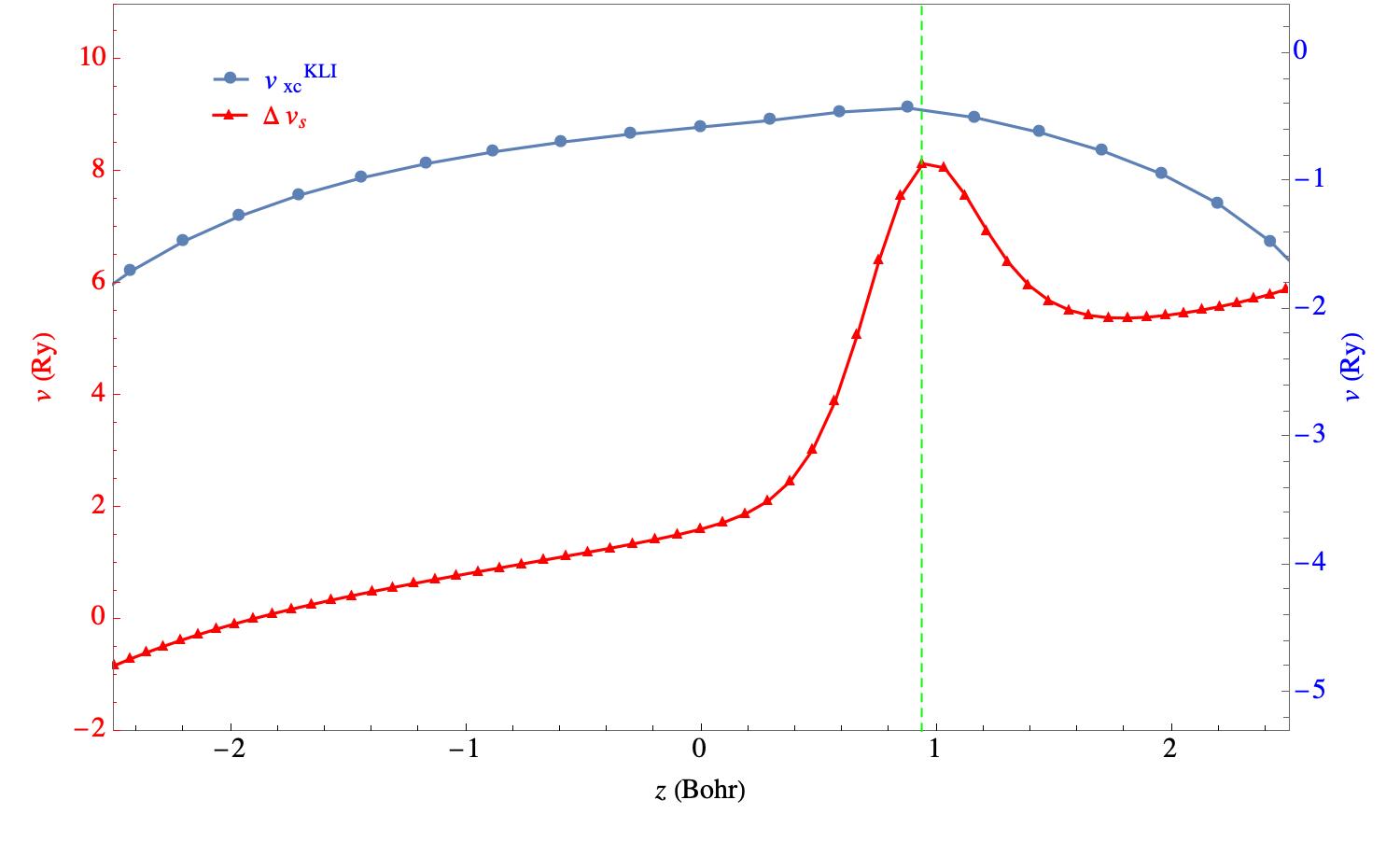}};
   \node [anchor=north west] (imgC) at (-0.275\linewidth,.35\linewidth)
            {\includegraphics[width=0.8\linewidth]{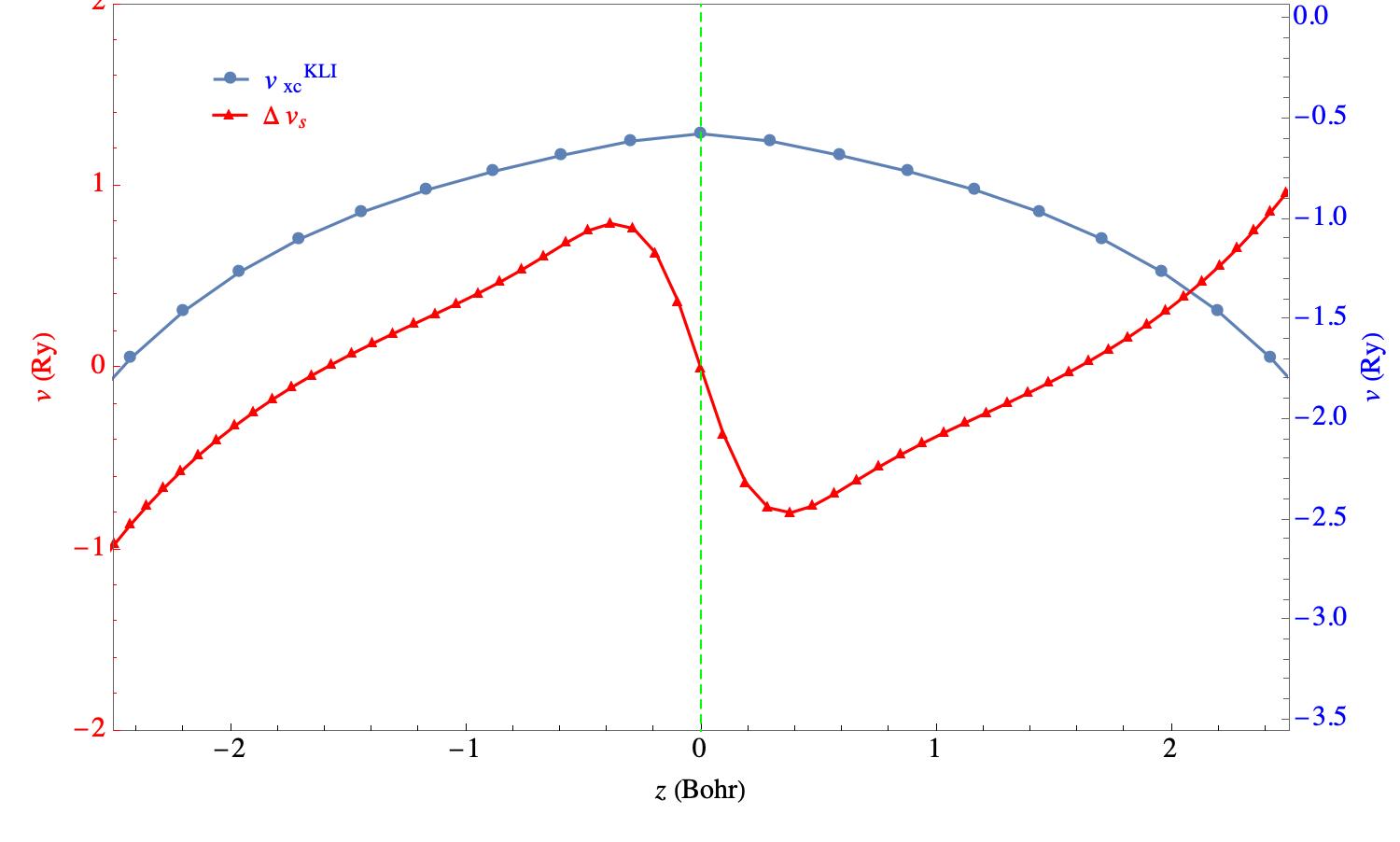}};
    \draw [anchor=north west] (-0.2\linewidth, .95\linewidth) node {(a) {\fontfamily{Arial}\selectfont {}}};
    \draw [anchor=north west] (-0.2\linewidth, .41\linewidth) node {(b) {\fontfamily{Arial}\selectfont {}}};
\end{tikzpicture}
\captionsetup{justification=raggedright,singlelinecheck=false}
\caption{Deducing the position of the step  from the difference between $v_s[\rho_A]$ and $v_s[\rho_B]$; a)  System: HeLi$^+$ where He is at $(0,0,-3.5)$ Bohr and Li is at $(0,0,3.5)$ Bohr; Red line: $\Delta v(\textbf{r})=v_s[\rho_A](\textbf{r})-v_s[\rho_B](\textbf{r})$; Blue Line:  $v^{\text{OEP-KLI}}_{xc}(\textbf{r})$; Vertical green line: the location of the Step.
b) System:He-Hew here He is at $(0,0,-3.5)$ Bohr and He is at $(0,0,3.5)$ Bohr; Red line: $\Delta v(\textbf{r})=v_s[\rho_A](\textbf{r})-v_s[\rho_B](\textbf{r})$; Blue Line:  $v^{\text{OEP-KLI}}_{xc}(\textbf{r})$; Vertical green line: the location of the Step.}
\label{StepDeltaVsAnalyeLi}
 \end{figure}

The difference between charge densities $\Delta v(\textbf{r})$ is plotted in Fig.[\ref{StepDeltaVsAnalyeLi}] and compared to the $v^{\text{OEP-KLI}}_{xc}(\textbf{r})$. Although OEP being locally approximated by the KLI theory is highly accurate, it might vary slightly from the exact $v_{xc}(\textbf{r})$. After all, the $v^{\text{OEP-KLI}}_{xc}(\textbf{r})$ is the best available candidate to evaluate the SS of the $v^{\text{NAD}}[\rho_B,\rho_{tot}(\textbf{r})]$.

	The position of the step shown in Fig.[\ref{StepDeltaVsAnalyeLi}.a] from two theories, although are infinitesimally dis-matched (about 0.04 Bohr) but is a good confirmation of the accuracy of the $v^{\text{NAD}}[\rho_B,\rho_{tot}](\textbf{r})$.
In Fig.[\ref{StepDeltaVsAnalyeLi}.b], the results are shown for the He-He model system, and this time, both theory shows exactly the same position for the SS in both curves. 
\subsection{LDA vs KLI and $v^{\text{NAD}}[\rho_B,\rho^{\text{LDA}}_{tot}]$} 
The SS, as it was already mentioned, is one of the properties of the exact potential. Now that the $v^{\text{NAD}}[\rho_B,\rho_{tot}(\textbf{r})]$ provides precisely this information, it could be used as a reference for the evaluation of the other theories. 

\begin{figure}[h]
\begin{tikzpicture}
	\node [anchor=north west] (imgA) at (-0.275\linewidth,.90\linewidth)
			{\includegraphics[width=0.8\linewidth]{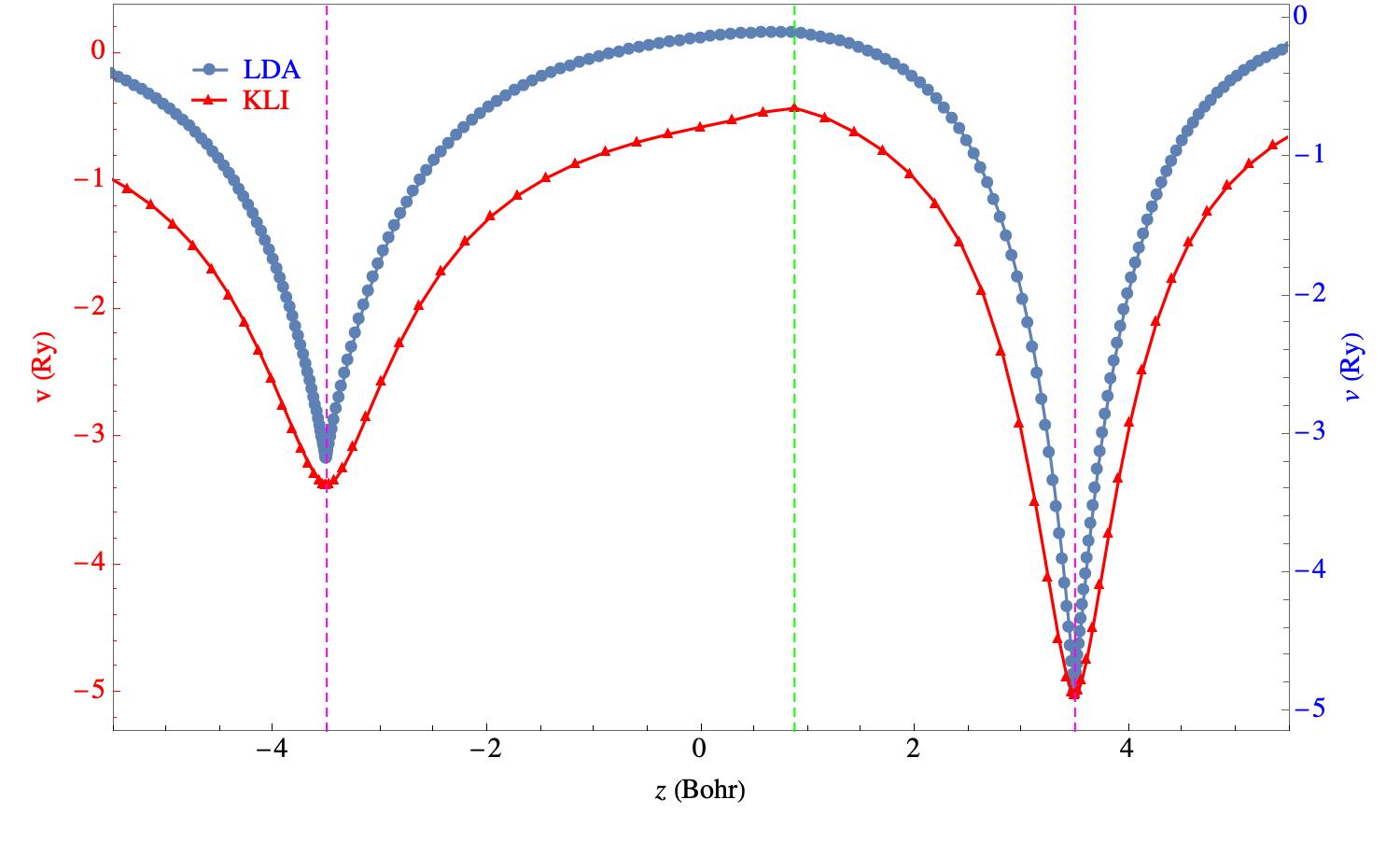}};
   \node [anchor=north west] (imgC) at (-0.275\linewidth,.35\linewidth)
            {\includegraphics[width=0.8\linewidth]{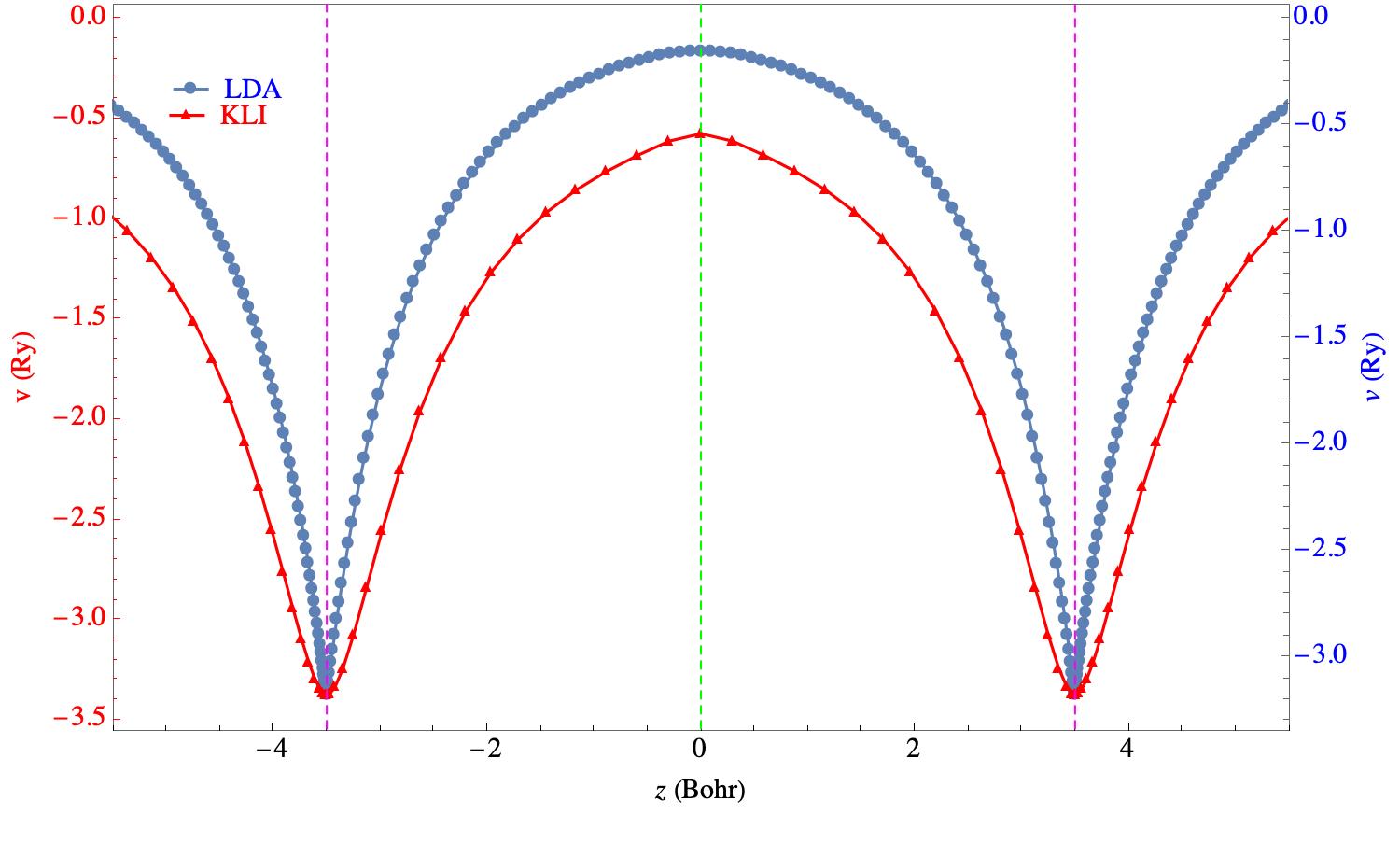}};
    \draw [anchor=north west] (-0.2\linewidth, .95\linewidth) node {(a) {\fontfamily{Arial}\selectfont {}}};
    \draw [anchor=north west] (-0.2\linewidth, .41\linewidth) node {(b) {\fontfamily{Arial}\selectfont {}}};
\end{tikzpicture}
\captionsetup{justification=raggedright,singlelinecheck=false}
\caption{Assessment of $v^{\text{LDA}}_{xc}(\textbf{r})$ from the appearance or nonappearance of the Step on its curve vs the $v^{\text{OEP-KLI}}_{xc}(\textbf{r})$;
a)  System: HeLi$^+$ where He is at $(0,0,-3.5)$ Bohr and Li is at $(0,0,3.5)$ Bohr; Red line: $v^{\text{OEP-KLI}}_{xc}(\textbf{r})$; Blue Line:  $v^{\text{LDA}}_{xc}(\textbf{r})$; Vertical green line: the location of the Step.
b) System: He-He where He is at $(0,0,-3.5)$ Bohr and He is at $(0,0,3.5)$ Bohr; Red line: $v^{\text{OEP-KLI}}_{xc}(\textbf{r})$; Blue Line:  $v^{\text{LDA}}_{xc}(\textbf{r})$; Vertical green line: the location of the Step.}
\label{StepLDAandKLIHeLi}
 \end{figure}

The LDA was already evaluated to be not exact but accurate enough to be used as a low-cost theory within the Kohn-Sham formulation of the DFT. We chose this theory to show that it misses the physical information carried by SS on the $v_{xc}(\textbf{r})$ curve. 
The $v^{\text{LDA}}_{xc}(\textbf{r})$ is compared with the $v^{\text{OEP-KLI}}_{xc}(\textbf{r})$ and the $v^{\text{NAD}}[\rho_B,\rho^{\text{LDA}}_{tot}]$ for He-He and HeLi$^+$ in Fig.[\ref{StepLDAandKLIHeLi}].
\subsection{Energy Gap}\label{GapFromVNAD}
The energy gap is the first physical property to be deduced from the SS. 
In a section \ref{OSS}, we explained that the origin of the step is related to the atomic $\epsilon^{HO/LU}$ or the molecular $\eta^{HO/LU}_{R/L}$ (see Eq.[\ref{StepHeight1}] and Eq.[\ref{StepDelta}]).

Within a heteronuclear diatomic close-shell system, the step height is related to the difference between HOMO and LUMO of the whole system but mathematically can be deduced from both the localised charge densities and the $v^{\text{NAD}}[\rho_B,\rho^{\text{LDA}}_{tot}](\textbf{r})$.

When it concerns the charge distribution of the sub-densities, the position of the step has to match the local extremum of the charge density that represents the density around the larger nuclei (here $\rho_A(\textbf{r})$ that is formed around atom Li).

However, when the system concerns a homonuclear diatomic case, the step is expected to occur in the middle of the interatomic distance, and the position of the step has to match the intersection of two sub-densities' curves.
\begin{figure}[h]
\begin{tikzpicture}
	\node [anchor=north west] (imgA) at (-0.4\linewidth,.90\linewidth)
			{\includegraphics[width=0.8\linewidth]{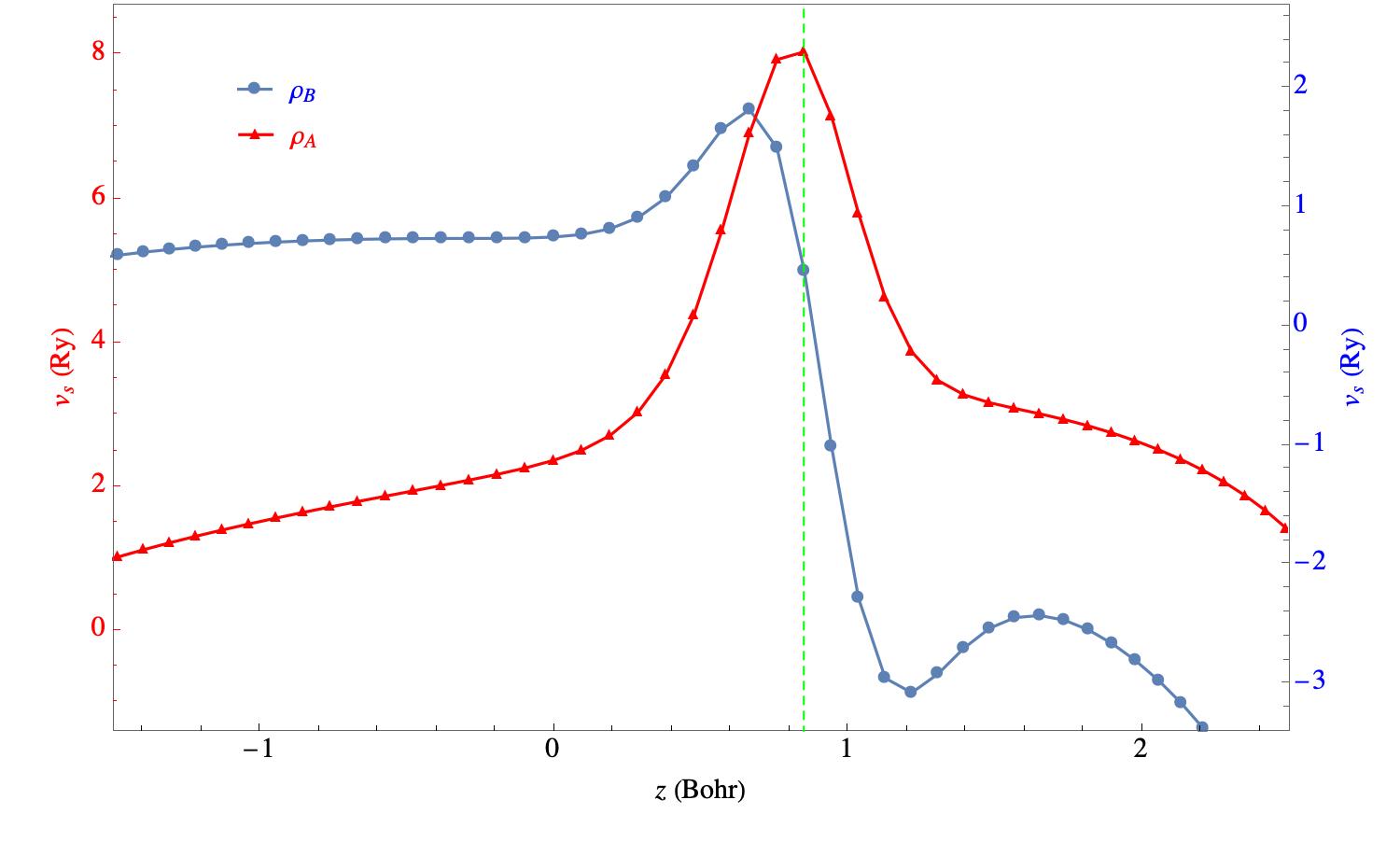}};
   \node [anchor=north west] (imgC) at (-0.4\linewidth,.35\linewidth)
            {\includegraphics[width=0.8\linewidth]{Graphs/Steps/HeHeVxcLDAvsKLI.jpg}};
    \draw [anchor=north west] (-0.15\linewidth, .95\linewidth) node {(a) {\fontfamily{Arial}\selectfont {}}};
    \draw [anchor=north west] (-0.15\linewidth, .41\linewidth) node {(b) {\fontfamily{Arial}\selectfont {}}};
\end{tikzpicture}
\captionsetup{justification=raggedright,singlelinecheck=false}
\caption{Analytical inversion of the potential from $\rho_A(\textbf{r})$ (localised around Li at the right side of the interatomic axis) and $\rho_B(\textbf{r})$ (localised around He at the left side of the interatomic axis); 
a) HeLi$^+$, where He is at $(0,0,-3.5)$ Bohr and Li, is at $(0,0,3.5)$ Bohr; Red line: $v^{\text{EXACT}}_s[\rho_A](\textbf{r})$; Blue line: $v^{\text{EXACT}}_s[\rho_B](\textbf{r})$; Vertical green line: the location of the Step observed on the $v^{\text{NAD}}[\rho_B,\rho_{tot}(\textbf{r})]$.
b) System: He-He where He is at $(0,0,-3.5)$ Bohr and Li is at $(0,0,3.5)$ Bohr; Red line: $v^{\text{EXACT}}_s[\rho_A](\textbf{r})$; Blue line: $v^{\text{EXACT}}_s[\rho_B](\textbf{r})$; Vertical green line: the location of the Step observed on the $v^{\text{NAD}}[\rho_B,\rho_{tot}(\textbf{r})]$.}
\label{StepVsAnalyHeHe}
 \end{figure}

Fig.[\ref{StepVsAnalyHeHe}] shows the position of the step happening accurately at the intersection of the two graphs.

\begin{figure}[ht]
	\centering
		\includegraphics[width=9cm]{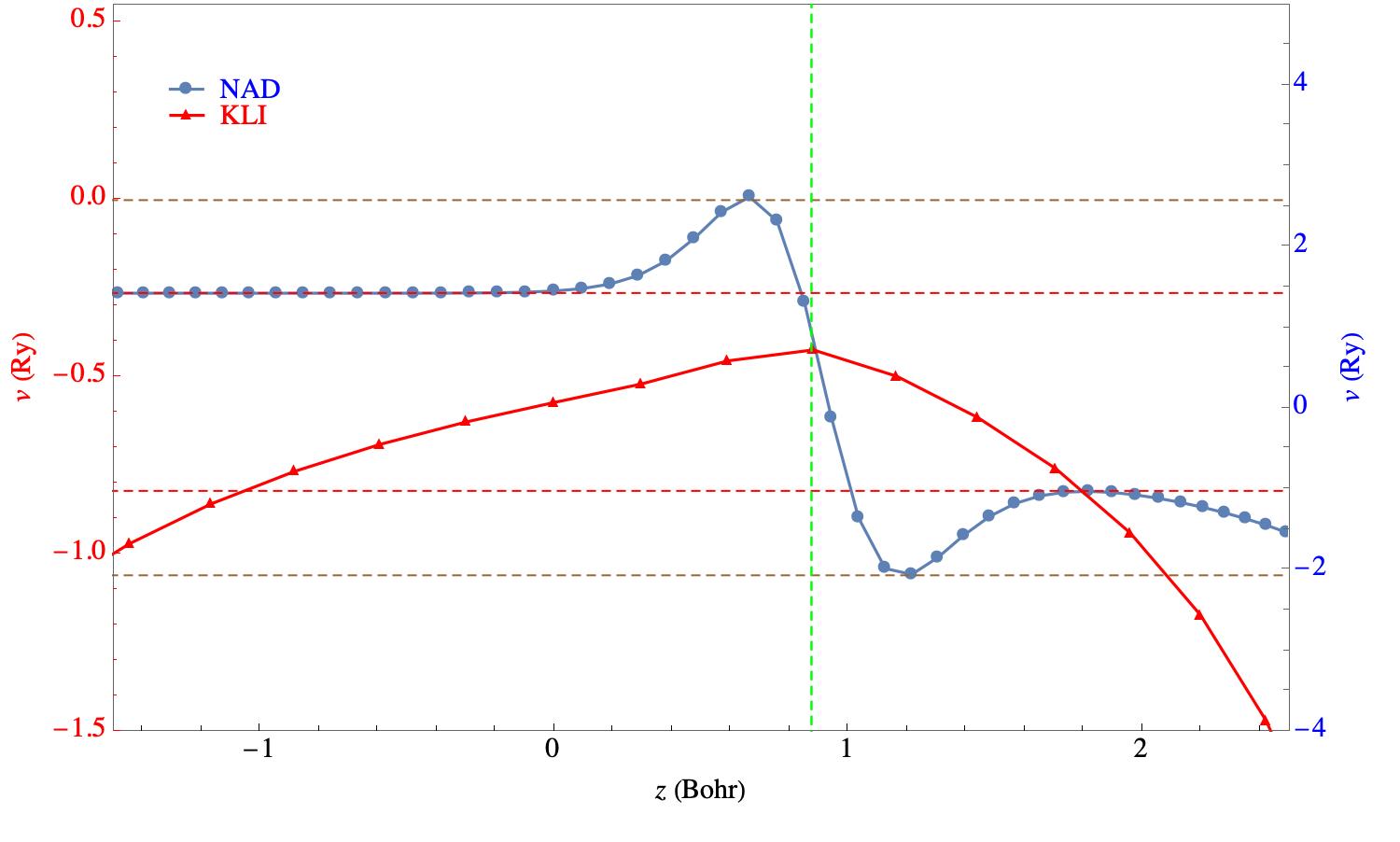}
  \captionsetup{justification=raggedright,singlelinecheck=false}
		\caption{ Finding the energy gap from the exact  $v^{\text{NAD}}[\rho_B,\rho_{tot}](\textbf{r})$; System: HeLi$^+$ where He is at $(0,0,-3.5)$ Bohr and Li is at $(0,0,3.5)$ Bohr; Vertical green line: location of the step. The $E_{g}$ is the half value of the difference between the tangent to the inflection points (horizontal brown lines) before the local extremum of the cure at the neighbouring step's position (horizontal red lines). }
		\label{GapHeLiVNADvsKLI}
	\end{figure}

The tangents to the second spatial derivative of the $v^{\text{NAD}}[\rho_B,\rho_{tot}](\textbf{r})$ (We mean $\frac{d^2v^{\text{NAD}}[\rho_B,\rho_{tot}](\textbf{r})}{d\textbf{r}^2}$) at the vicinity of the step position have been parallel to the interatomic axis. The half of the difference between the energy related to each tangent is the $E_g$. If that is the correct value, then the exact position of the step must occur spatially exactly at the mid-distance of two local extrema. 
If Fig.[\ref{GapHeLiVNADvsKLI}] The vertical green line shows where the step happened on $v^{\text{OEP-KLI}}_{xc}(\textbf{r})$. We see that this line is not happening at the mid-distance of two local extrema of the exact potential. That means the $E^{\text{EXACT}}_g$ must be underestimated compared to the $E^{\text{NAD}}_g$ for when the localised density is highly comparable with any of the $|\phi_i|^2(\textbf{r})$. 
In Table [\ref{tableEgsLiHe}] we see clearly this underestimation of the $E^{\text{EXACT}}_g$ compared to the $E^{\text{NAD}}_g$ by a value about $1.6 \times 10^{-3}$ Ry. 
To have a more accurate conclusion from this comparison, we also want to see how the energy gap calculated by $v^{\text{NAD}}$ varies for non-localised charge density.

\subsection{Non-localised Electrons}
This comparison between the $E^{\text{EXACT}}_g$ and the $E^{\text{NAD}}_g$ could be not justified if both theories don't share the same density distribution of sub-densities in order to calculate the $E_g$. 

In fact, the $v^{\text{OEP-KLI}}_{KS}$ was suggested to solve accurately a molecular system in which the molecular orbital densities could partially surpass the integer value while being integrated spatially in the whole space.
So, it is more justified if we evaluate the $OEP-KLI$ approach with comparing its $E^{\text{EXACT}}_g$ with energy gap obtained from $v^{\text{NAD}}[\rho_{B'},\rho^{\text{LDA}}_{tot}](\textbf{r})$ where $\int \rho_{B'}(\textbf{r})d\textbf{r}=2\pm \delta$.


\begin{figure}[h]
\begin{tikzpicture}
	\node [anchor=north west] (imgA) at (-0.275\linewidth,.90\linewidth)
			{\includegraphics[width=0.8\linewidth]{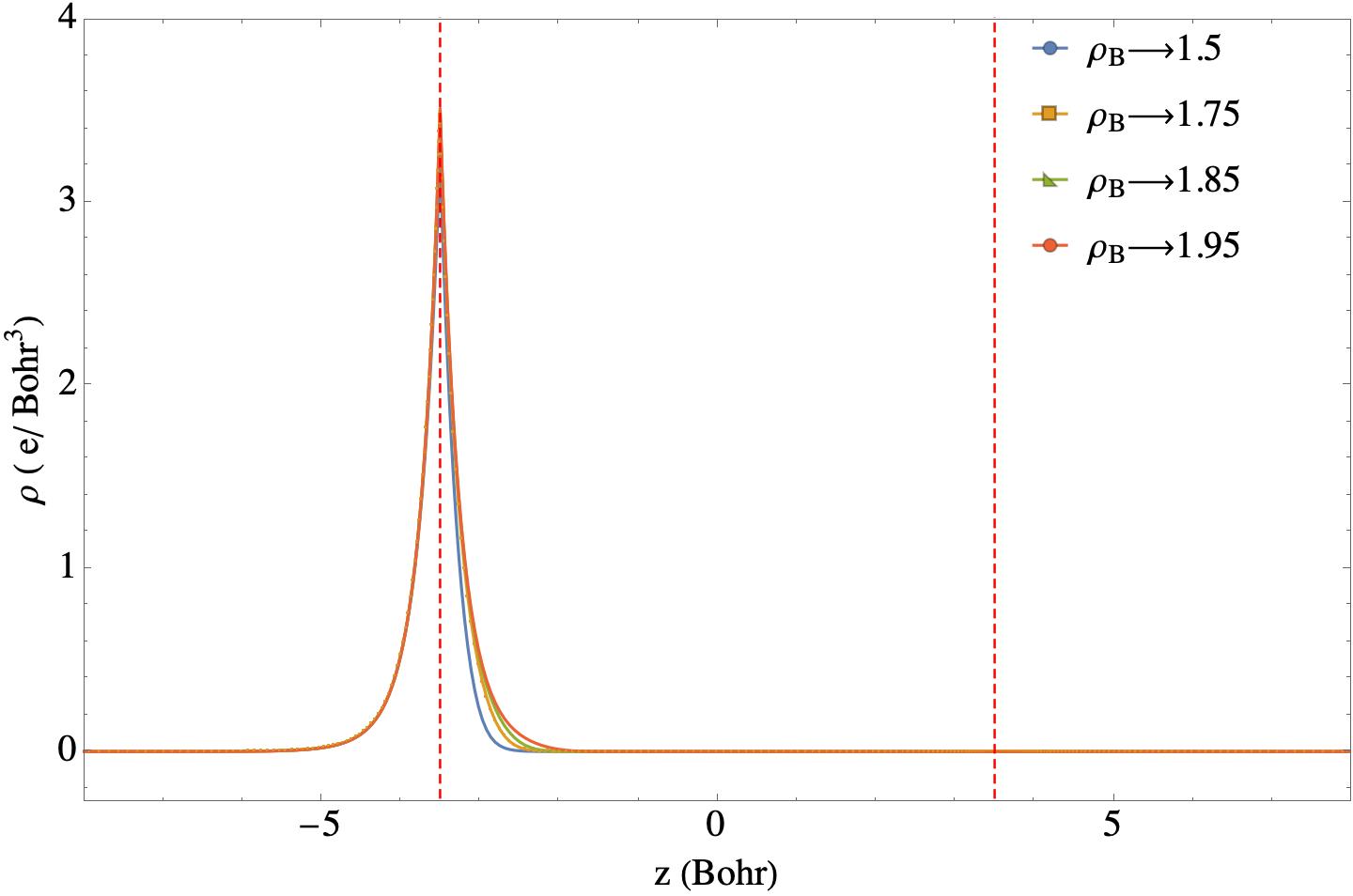}};
   \node [anchor=north west] (imgC) at (-0.275\linewidth,.35\linewidth)
            {\includegraphics[width=0.8\linewidth]{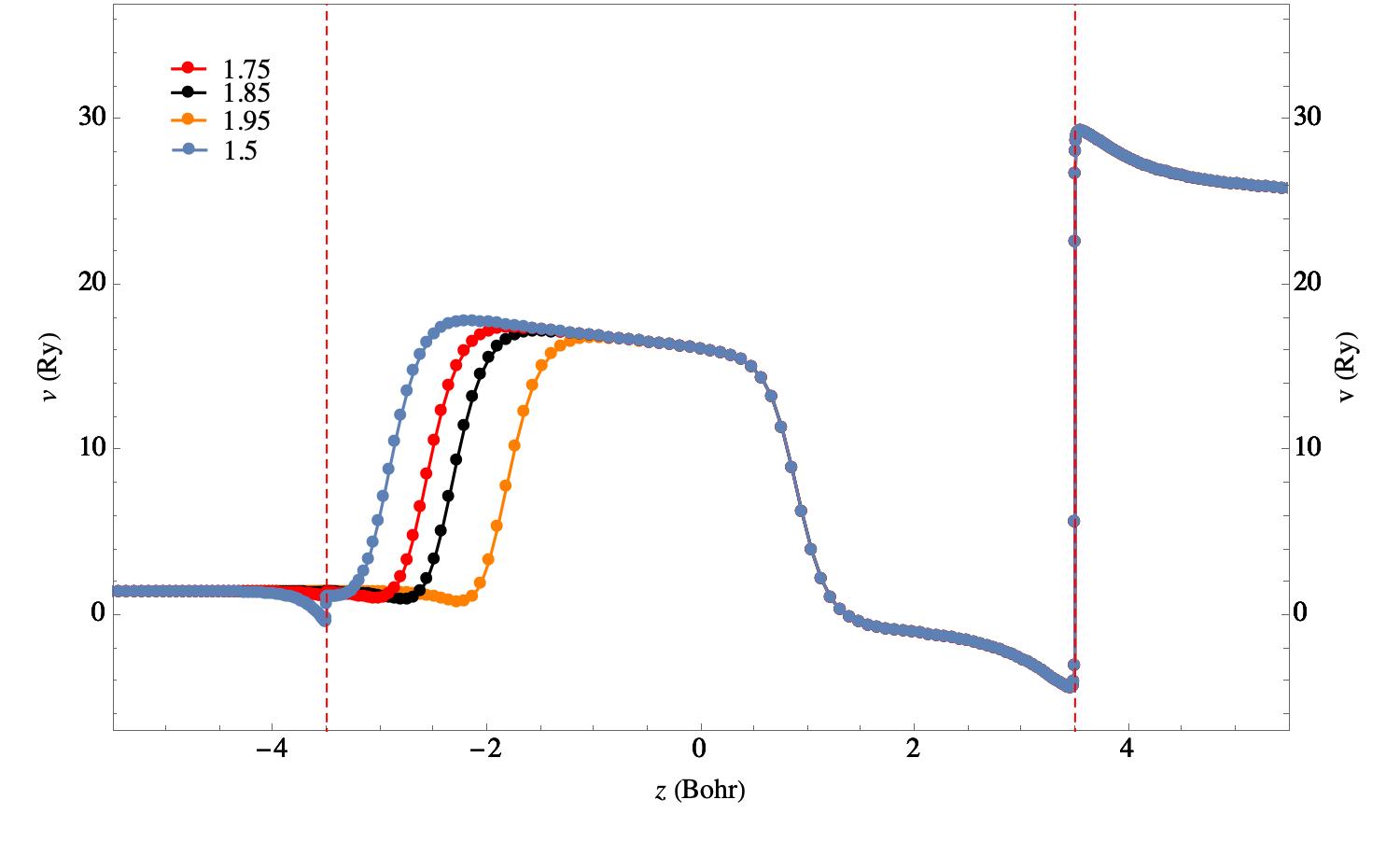}};
    \draw [anchor=north west] (-0.2\linewidth, .95\linewidth) node {(a) {\fontfamily{Arial}\selectfont {}}};
    \draw [anchor=north west] (-0.2\linewidth, .41\linewidth) node {(b) {\fontfamily{Arial}\selectfont {}}};
\end{tikzpicture}
\captionsetup{justification=raggedright,singlelinecheck=false}
 \caption{  System: HeLi$^+$ where He is at $(0,0,-3.5)$ Bohr and Li is at $(0,0,3.5)$ Bohr; Vertical brown line: nuclei; 
 (a) Different partial localisation of the charge density around the He nuclei at the left side of the interatomic axis;
  Total and partitioned densities. 
 (b)The exact $v^{\text{NAD}}[\rho_B,\rho_{tot}](\textbf{r})$ from analytical inversion for different partial localisation of the charge density around the He nuclei at the left side of the interatomic axis plotted in figure above; The $E_{g}\simeq 1.44$ (Ry).
 }
\label{1dHeHechgDensInvvW}
 \end{figure}

\subsection{Heteronuclear System}
\subsection{HeLi$^+$}
The $v^{\text{NAD}}[\rho_{B'},\rho_{tot}](\textbf{r})$ for partially localised charge density shown in Fig.[\ref{1dHeHechgDensInvvW}.b] compared to inverted potential from the integer localisation of charge density (Fig.[\ref{StepKLIHeLi}.a]) shows a significant change on the potential curve. 
The change of the future of the potential inverted from partial localisation, although it appears on the overlap region and close to the nuclei around which the charge density is localised remains comparable for different partially localised densities.

The jump of the step in between two atoms has a similar height for different $\rho_B(\textbf{r})$s. However, the closer the charge localisation is to the integer value, the more the jump (from the left to the right) on the potential is shifted to the right side, and the smaller gets the width of the step. 

As we asserted previously that the exact position of the Step must occur at the local extremum of the inverted potential from the reminder charge density, it's essential to verify this fact by looking at the $v_s[\rho_i](\textbf{r})$ for $i=A, B$.

\begin{figure}[ht]
	\centering
		\includegraphics[width=9cm]{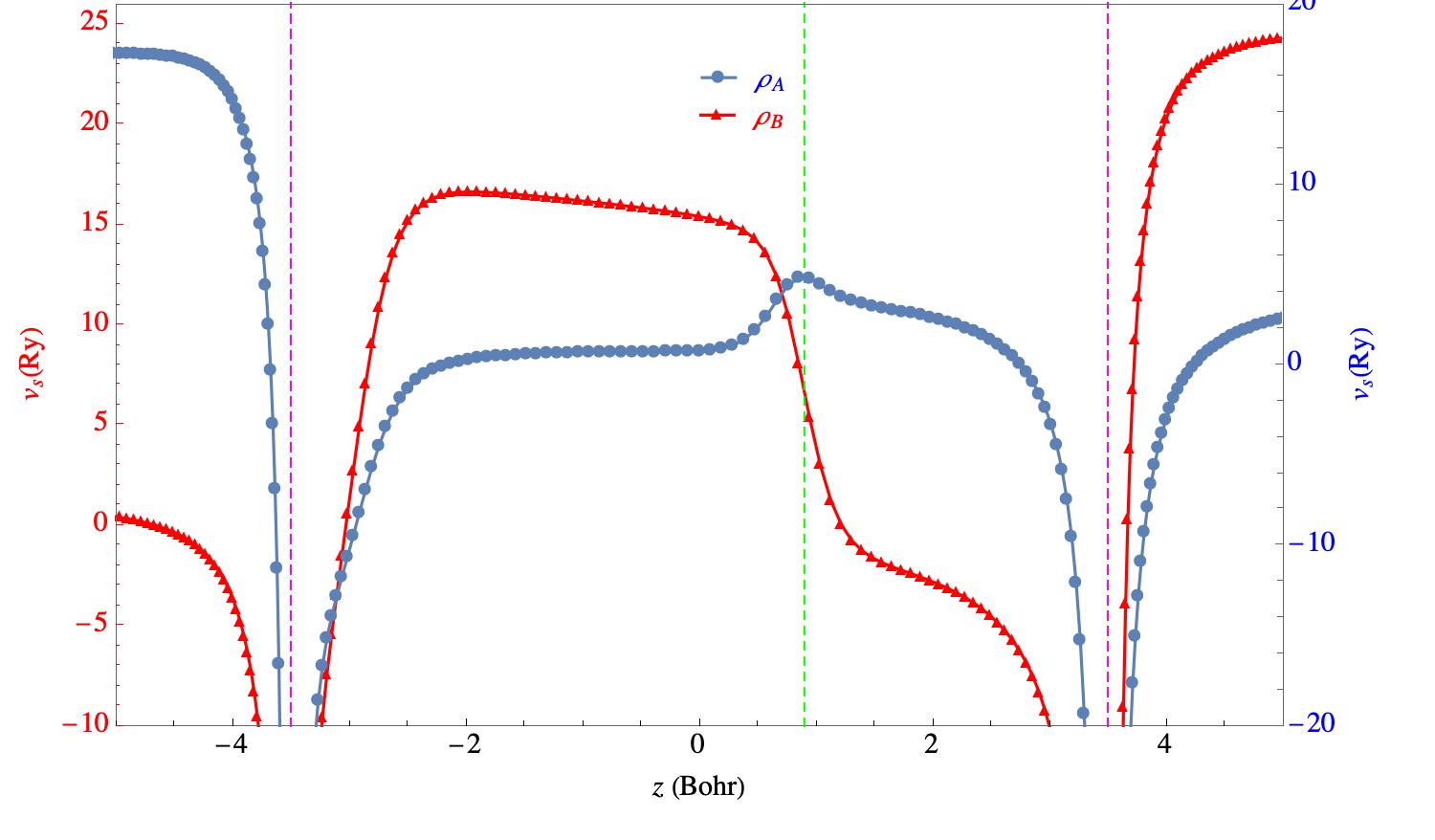}
  \captionsetup{justification=raggedright,singlelinecheck=false}
		\caption{Analytical inversion of the potential from $\rho_A(\textbf{r})$ (localised around Li at the right side of the interatomic axis) and $\rho_B(\textbf{r})$ (localised around He at the left side of the interatomic axis) for partial localisation ($\int \rho_B(\textbf{r}) d\textbf{r}=1.5$) of the charge density around the He nuclei at the left side of the interatomic axis plotted in Fig.[\ref{1dHeHechgDensInvvW}. a]; System: HeLi$^+$ where He is at $(0,0,-3.5)$ Bohr and Li is at $(0,0,3.5)$ Bohr; Vertical brown line: nuclei. The $E_{g}\simeq 1.44$ (Ry).}
		\label{LiHeVsAandB15zPlot}
	\end{figure}

In Fig.[\ref{LiHeVsAandB15zPlot}] the position of the step happens precisely at the bump of the $v_s[\rho_A](\textbf{r})$ and accurately crosses the mid-hight of the change in $v_s[\rho_B](\textbf{r})$. 

The $E^{\text{NAD}}_g$ this time is the same as the $E^{\text{EXACT}}_g$ (last line in Table.[\ref{tableEgsLiHe}]).
\begin{table}[ht]
\centering
\begin{tabular}{ |c    ||c|c|c|c| } 
\hline
Theory & \textbf{\rm HOMO} (Ry) & \textbf{\rm LUMO} (Ry) & \textbf{Gap} (Ry) \\
\hline\hline
\multirow{1}{4em}{\textbf{LDA}} & $ -2.814$ 
& $-0.415$ & $1.088$ \\ 
\hline
\multirow{1}{4em}{\textbf{KLI}} & $ -2.122$ 
& $-0.683$  & $1.439$ \\ 
\hline
\multirow{1}{4em}{\textbf{NAD}} & $--$ 
&  $--$   & $1.215$ \\
\hline
\multirow{1}{4em}{\textbf{NAD'}}   &   $--$ 
&  $--$   & \color{red}{$1.441$} \\
\hline
\end{tabular}
\captionsetup{justification=raggedright,singlelinecheck=false}
\caption{ The energy gap from different theories for the model system LiHe$^+$. NAD refers to the potential inverted from the density that integrates to 2, and NAD' corresponds to the potential inverted from a density that integrates to 1.5. All input densities are from LDA.}
\label{tableEgsLiHe}
\end{table}

	\begin{figure}[ht]
	\centering
		\includegraphics[width=9cm]{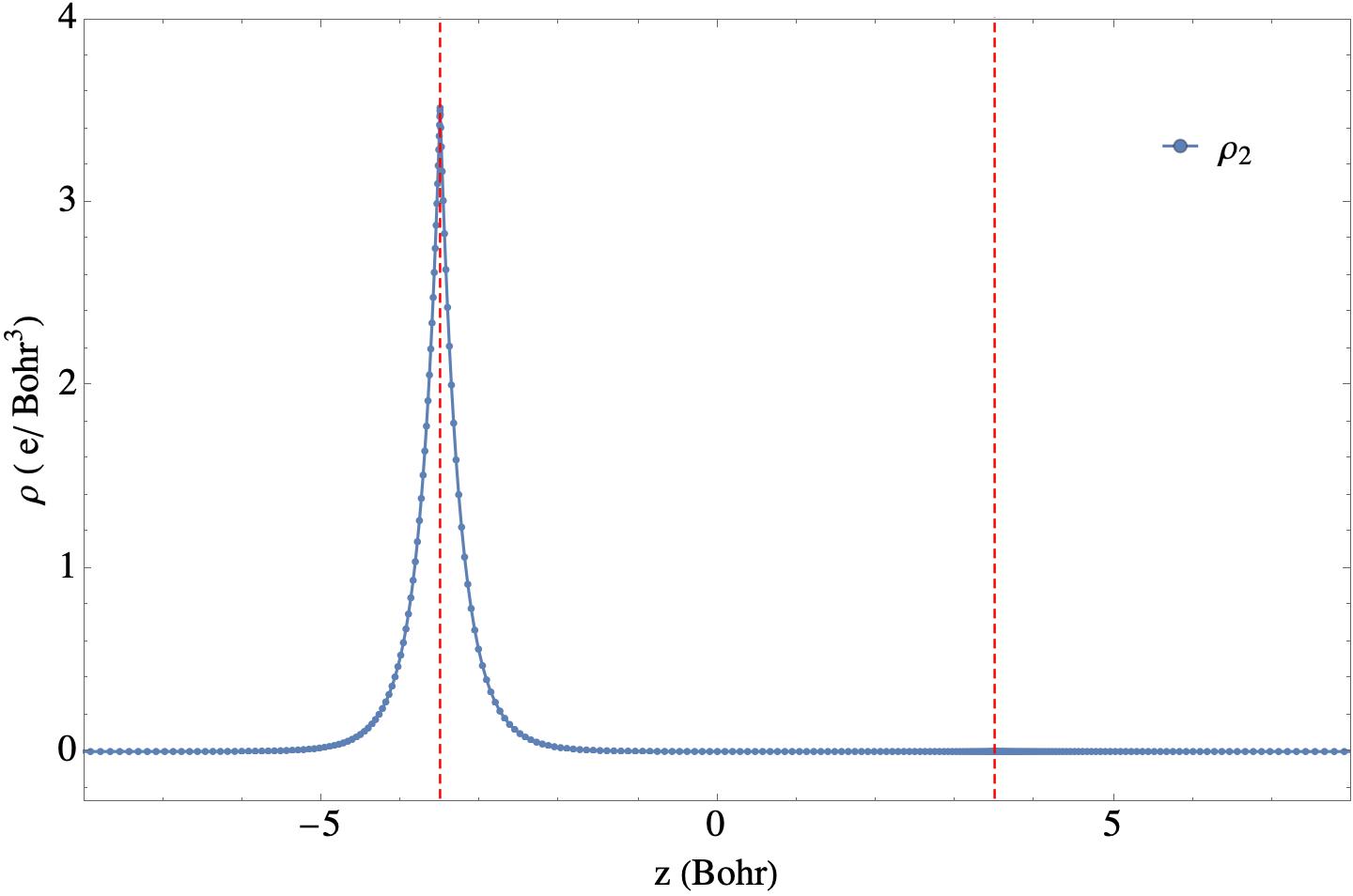}
  \captionsetup{justification=raggedright,singlelinecheck=false}
		\caption{1D representation of the $2s$ orbital $\rho_2(\textbf{r})=2|\phi_2(\textbf{r})|^2$ for the system HeLi$^+$; He is sited at $(0,0,-3.5)$ Bohr and Li is located at $(0,0,3.5)$ Bohr; Vertical brown line: nuclei.}
		\label{Rho2HeLi}
	\end{figure}

\begin{figure}[ht]
	\centering
		\includegraphics[width=9cm]{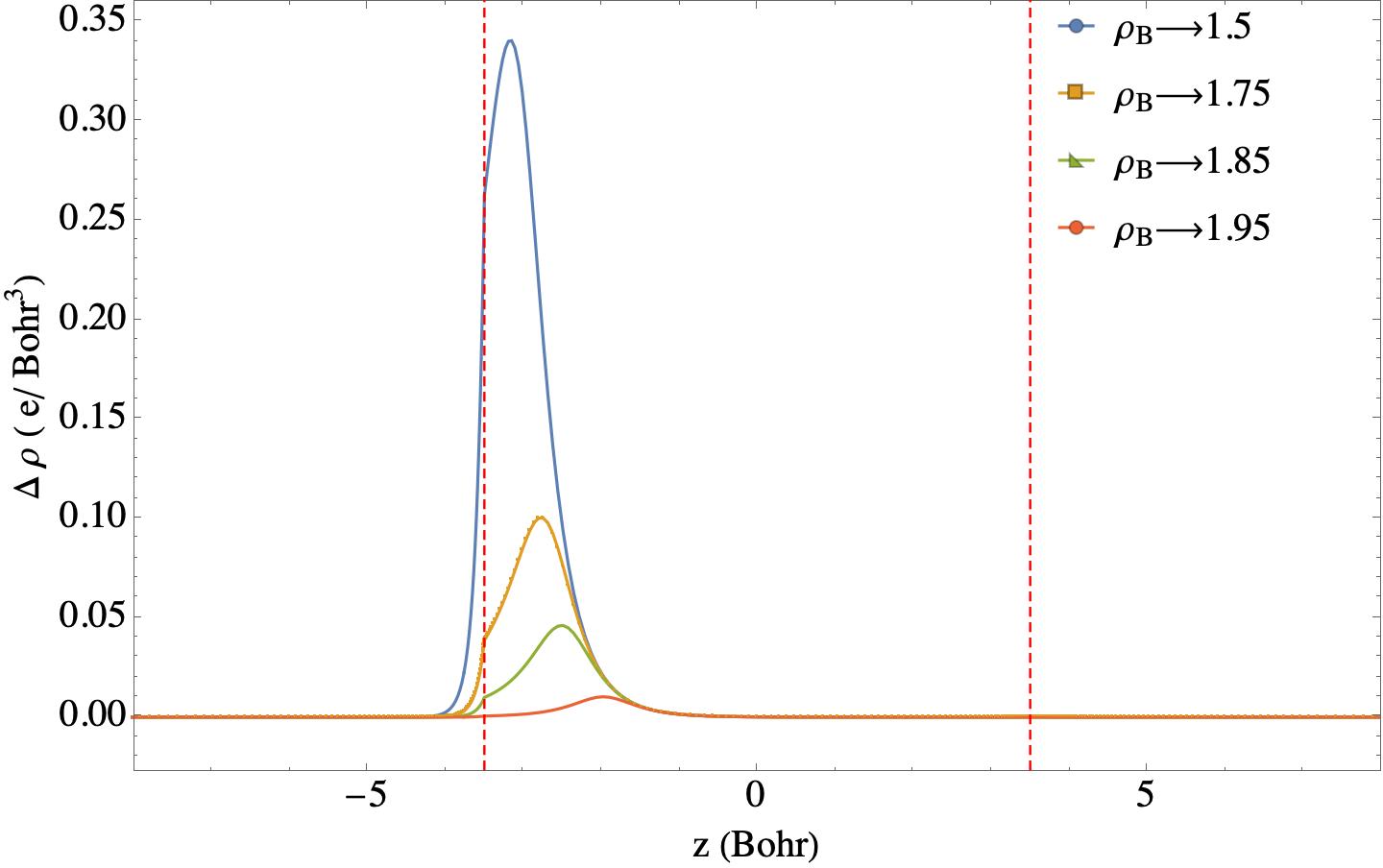}
  \captionsetup{justification=raggedright,singlelinecheck=false}
		\caption{Variation of the partially localised $\rho_B(\textbf{r})$ from the $2s$ orbital-density around the He nuclei; System: HeLi$^+$; He is sited at $(0,0,-3.5)$ Bohr and Li is located at $(0,0,3.5)$ Bohr; Vertical brown line: nuclei; coloured curves:  $\Delta\rho(\textbf{r})=\rho_2(\textbf{r})-\rho_B(\textbf{r})=\rho_1(\textbf{r})-\rho_A(\textbf{r})$ for $1.5 \leq \int \rho_B(\textbf{r})d\textbf{r}\leq1.95$.}
		\label{Deltarho2rhoBLiHeFrom15To195}
	\end{figure}

We plotted the $\rho_2(\textbf{r})=2|\phi_2(\textbf{r})|^2$ in Fig.[\ref{Rho2HeLi}] and showed the $\Delta\rho(\textbf{r})=\rho_2(\textbf{r})-\rho_B(\textbf{r})=\rho_1(\textbf{r})-\rho_A(\textbf{r})$ if Fig.[\ref{Deltarho2rhoBLiHeFrom15To195}] for different partially localised charge densities. 

The analytically inverted potential bi-functional needs to predict the missing charge to compensate $\rho_B(\textbf{r})$ to become a ground state orbital density. 
\subsection{HLi}
\begin{figure}[ht]
	\centering
		\includegraphics[width=9cm]{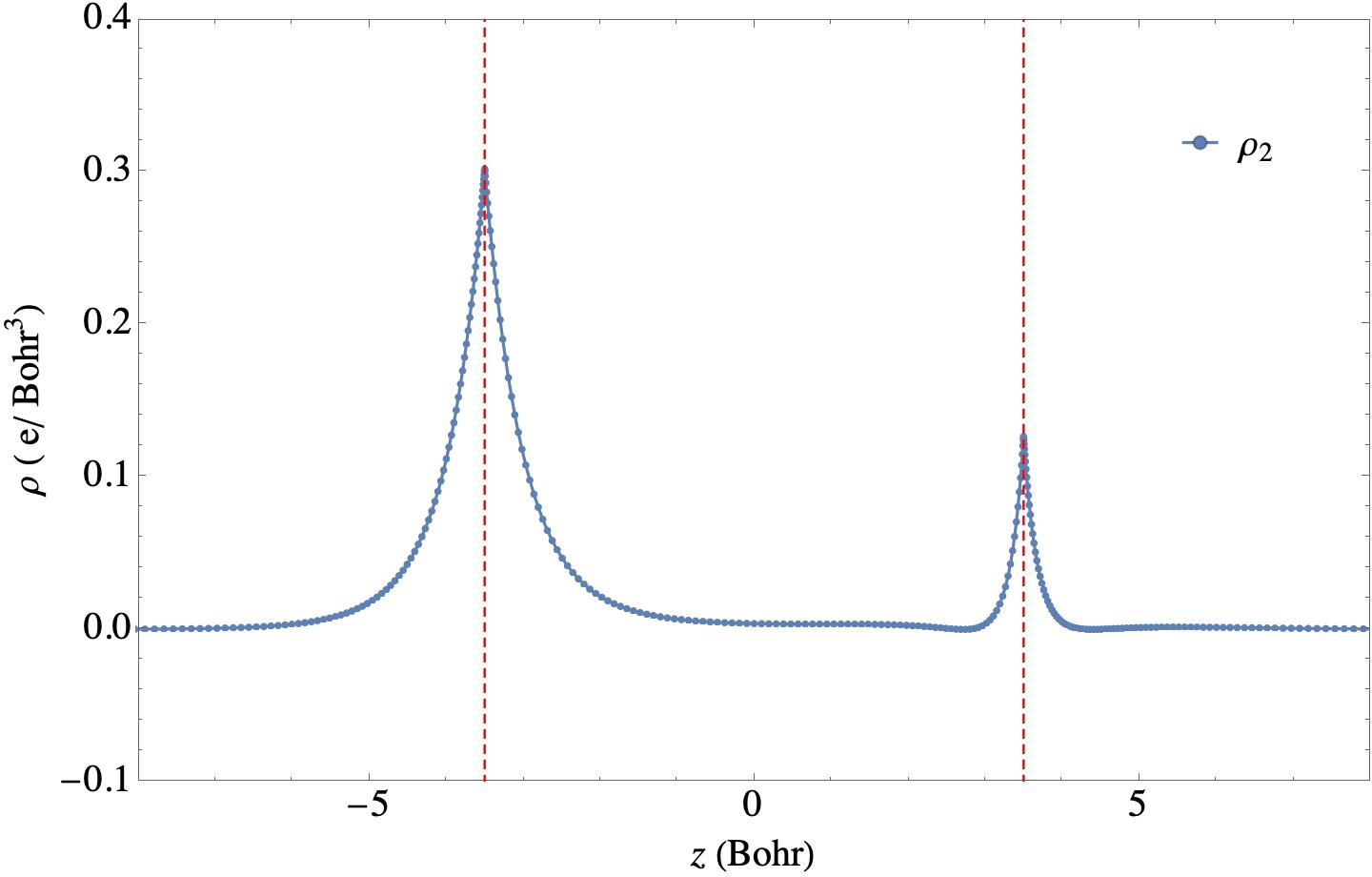}
\captionsetup{justification=raggedright,singlelinecheck=false}		
  \caption{1D representation of the Orbital density; System: HLi; H at $(0,0,-3.5)$ Bohr and Li at $(0,0,3.5)$ Bohr; the density is $\rho_2(\textbf{r})=2|\phi_2(\textbf{r})|^2$.}
		\label{rho2HLi2}
	\end{figure}

To verify the $v^{\text{NAD}}[\rho_{B'},\rho_{tot}](\textbf{r})$ for the more realistic system, the HLi molecule was chosen to be studied. First, we take a look at the first orbital density distribution in space to understand its locality (Fig.[\ref{rho2HLi2}]). 

\begin{figure}[ht]
	\centering
		\includegraphics[width=9cm]{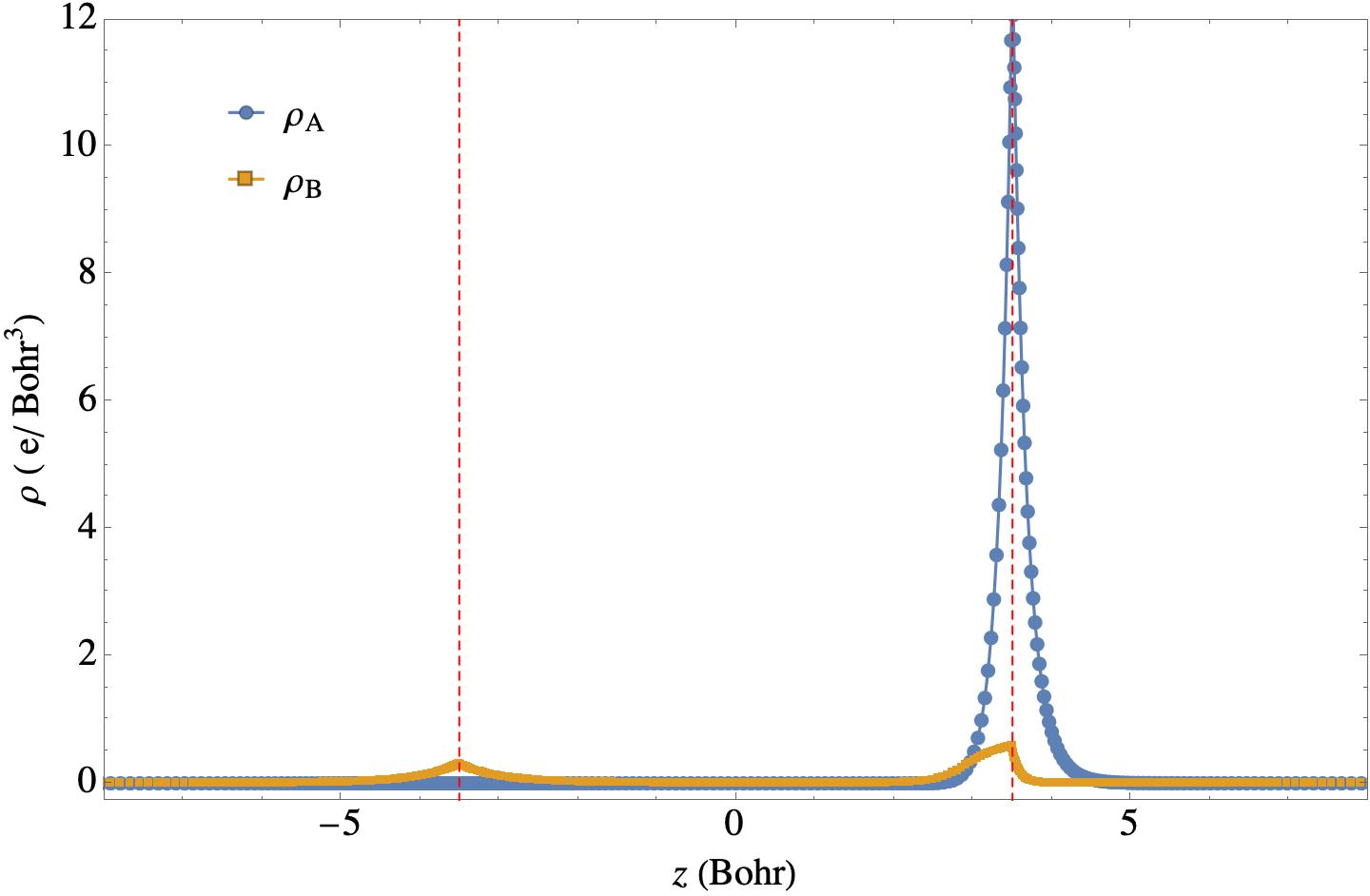}
  \captionsetup{justification=raggedright,singlelinecheck=false}
		\caption{1D representation of the localised density; System: HLi; H at $(0,0,-3.5)$ Bohr and Li at $(0,0,3.5)$ Bohr; Blue mark-line: $\rho_B(\textbf{r})$ as $\int_V\rho_B(\textbf{r})d\textbf{r}=2$; Orange mark-line: $\rho_A(\textbf{r})=\rho_{tot}(\textbf{r})-\rho_B(\textbf{r})$ }
		\label{rhosABHLi2}
	\end{figure}

Obviously, for HLi, the $\rho_B(\textbf{r})$ nor$\rho_A(\textbf{r}=\rho_{tot}(\textbf{r}-\rho_B(\textbf{r})$ for the condition in which $\int_V \rho_B(\textbf{r})d\textbf{r}=2$ cannot be localised at the vicinity of one of the atoms (Fig.[\ref{rhosABHLi2}]).
Instead, the $\rho_B(\textbf{r})$ become more localised around the H atom at the left side of the interatomic axis if $\int_V \rho_B(\textbf{r})d\textbf{r}=1.5$ (Fig.[\ref{rhosABHLi2}]).
\begin{figure}[h]
\begin{tikzpicture}
	\node [anchor=north west] (imgA) at (-0.275\linewidth,.90\linewidth)
			{\includegraphics[width=0.8\linewidth]{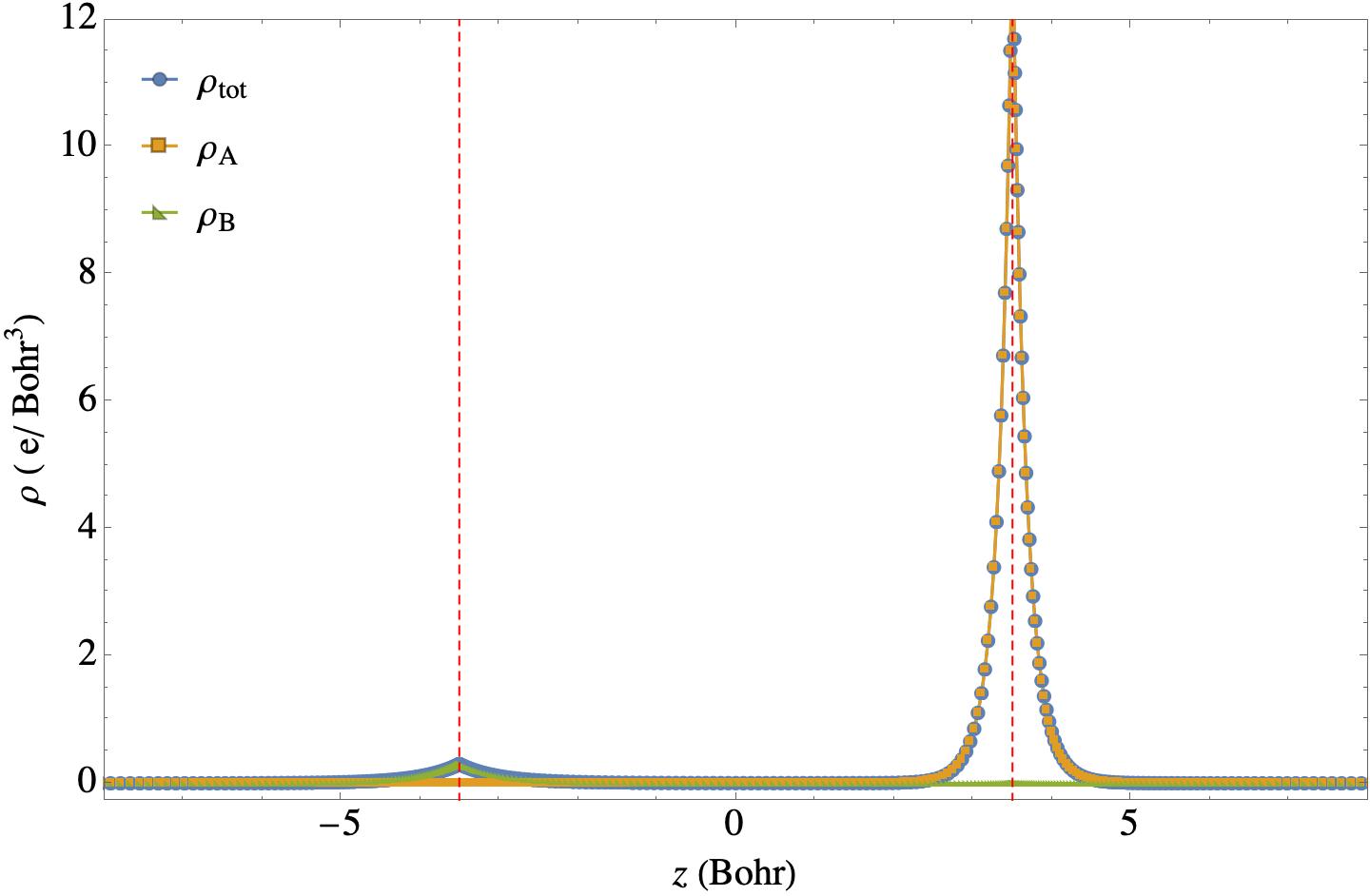}};
   \node [anchor=north west] (imgC) at (-0.275\linewidth,.35\linewidth)
            {\includegraphics[width=0.8\linewidth]{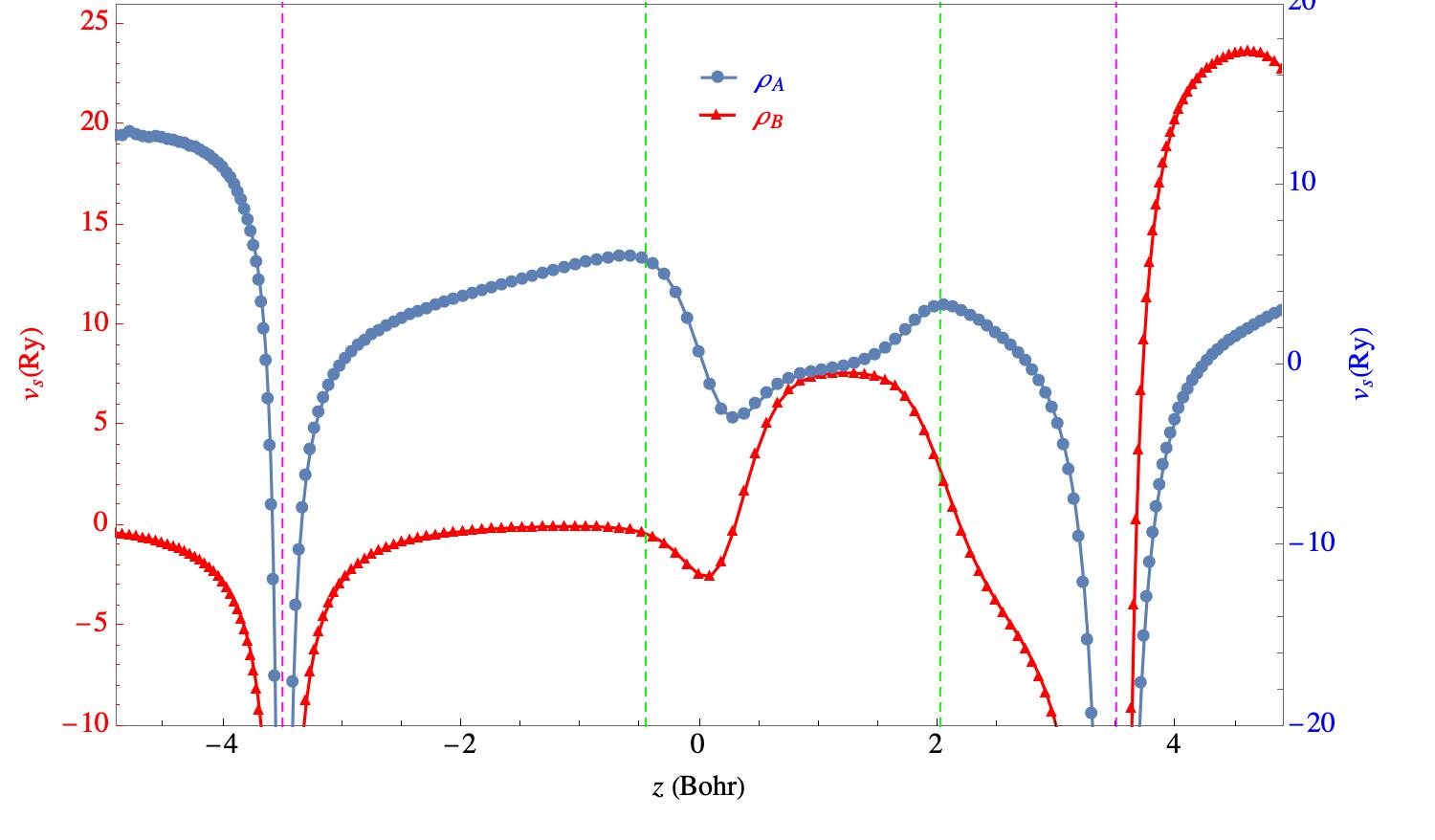}};
    \draw [anchor=north west] (-0.2\linewidth, .95\linewidth) node {(a) {\fontfamily{Arial}\selectfont {}}};
    \draw [anchor=north west] (-0.2\linewidth, .41\linewidth) node {(b) {\fontfamily{Arial}\selectfont {}}};
\end{tikzpicture}
 \captionsetup{justification=raggedright,singlelinecheck=false}
 \caption{
  System: HLi; H at $(0,0,-3.5)$ Bohr and Li at $(0,0,3.5)$ Bohr; Vertical brown line: nuclei;
 (a) 1D representation of the densities: 
 Blue mark-line: $\rho_{tot}(\textbf{r})$; Green mark-line: $\rho_B(\textbf{r})$ as $\int_V\rho_B(\textbf{r})d\textbf{r}=1.5$; Orange mark-line: $\rho_A(\textbf{r})=\rho_{tot}(\textbf{r})-\rho_B(\textbf{r})$; Vertical dashed lines: nuclei.
 (b)Analytical inversion of the potential from $\rho_A(\textbf{r})$ (localised around Li at the right side of the interatomic axis) and $\rho_B(\textbf{r})$ (localised around H at the left side of the interatomic axis) for partial localisation ($\int \rho_B(\textbf{r}) d\textbf{r}=1.5$) of the charge density around the He nuclei at the left side of the interatomic axis plotted in Fig.[\ref{1dHeHechgDensInvvW}.a]. 
 }
\label{HLiVsAandB15zPlot}
 \end{figure}

We choose the latter localisation of the $\rho_B(\textbf{r})$ and we provide the $v_s[\rho_A](\textbf{r})$ and $v_s[\rho_A](\textbf{r})$ (Fig.[\ref{HLiVsAandB15zPlot}]) from which we can obtain more information about the energy gap.

\begin{figure}[ht]
 \begin{tikzpicture}
	\node [anchor=north west] (imgA) at (-0.275\linewidth,.90\linewidth)
			{\includegraphics[width=0.8\linewidth]{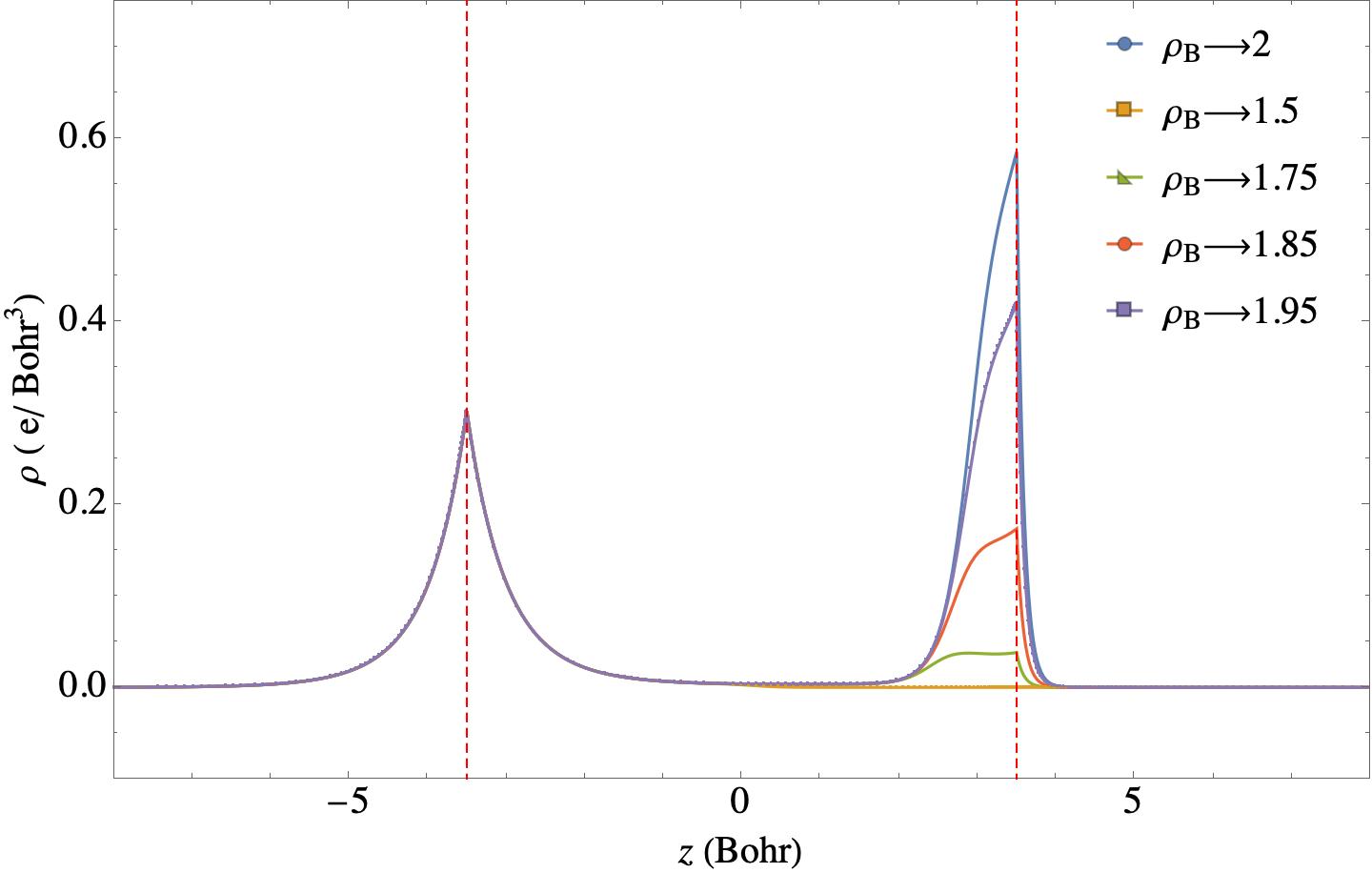}};
   \node [anchor=north west] (imgC) at (-0.275\linewidth,.35\linewidth)
            {\includegraphics[width=0.8\linewidth]{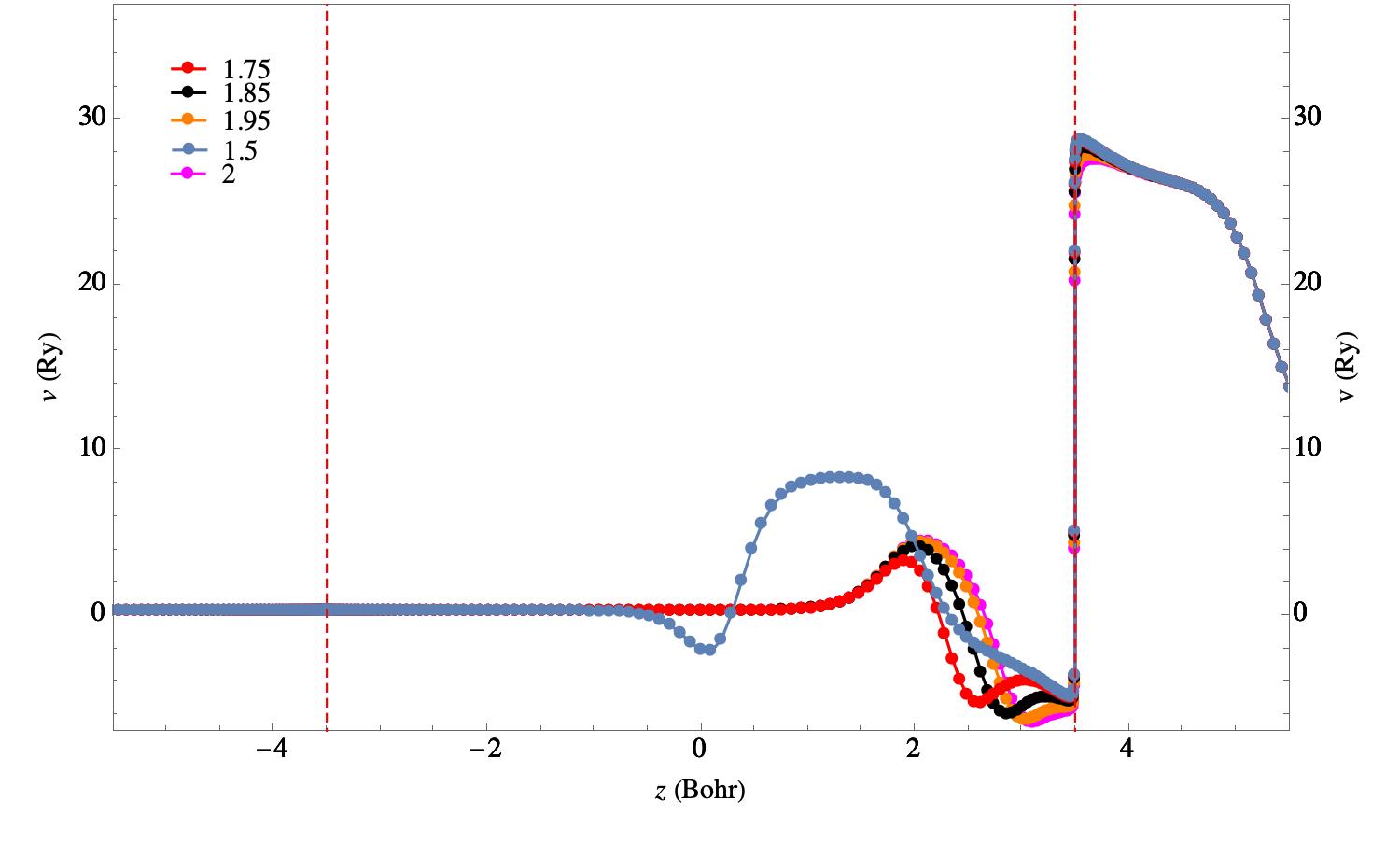}};
    \draw [anchor=north west] (-0.2\linewidth, .95\linewidth) node {(a) {\fontfamily{Arial}\selectfont {}}};
    \draw [anchor=north west] (-0.2\linewidth, .41\linewidth) node {(b) {\fontfamily{Arial}\selectfont {}}};
\end{tikzpicture}
	\captionsetup{justification=raggedright,singlelinecheck=false}
		\caption{ System: HLi; H is sited at $(0,0,-3.5)$ Bohr and Li is located at $(0,0,3.5)$ Bohr; Vertical brown line: nuclei;
  (a) Variation of the partially localised $\rho_B(\textbf{r})$ from the $2s$ orbital-density around the He nuclei; coloured curves:  $\Delta\rho(\textbf{r})=\rho_2(\textbf{r})-\rho_B(\textbf{r})=\rho_1(\textbf{r})-\rho_A(\textbf{r})$ for $1.5 \leq \int \rho_B(\textbf{r})d\textbf{r}\leq 2$.
  The exact $v^{\text{NAD}}[\rho_B,\rho_{tot}](\textbf{r})$ from analytical inversion for different partial localisation of the charge density around the He nuclei at the left side of the interatomic axis plotted in Fig.[\ref{1dHeHechgDensInvvW}.a].}
		\label{rhosHLiFrom15To195}
	\end{figure}

	As the $v^{\text{NAD}}[\rho_{B'},\rho_{tot}](\textbf{r})$ must carry the information about the difference between $\rho_B(\textbf{r})$ and $\rho_i\textbf{r})$ (for $i=1,2,3,...$), it's essential to know the feature of the $\Delta\rho(\textbf{r})=\rho_2(\textbf{r})-\rho_B(\textbf{r})=\rho_1(\textbf{r})-\rho_A(\textbf{r})$. For different density localisation around the H atom this difference is plotted in Fig.[\ref{rhosHLiFrom15To195}].
The corresponding $v^{\text{NAD}}[\rho_{B'},\rho_{tot}](\textbf{r})$ for different $\rho_B(\textbf{r})$ is demonstrated in 1D representation in Fig.[\ref{rhosHLiFrom15To195}].
	
\begin{table}[ht]
\centering
\begin{tabular}{ |c    ||c|c|c|c| } 
\hline
Theory & \textbf{\rm HOMO} (Ry) & \textbf{\rm LUMO} (Ry) & \textbf{Gap} (Ry) \\
\hline\hline
\multirow{1}{4em}{\textbf{LDA}} & $  -0.283$ 
& $ -0.111$ & $0.172$ \\ 
\hline
\multirow{1}{4em}{\textbf{NAD'}}   &   $--$ 
&  $--$   & \color{red}{$0.334$} \\
\hline

\end{tabular}
\captionsetup{justification=raggedright,singlelinecheck=false}
\caption{ The energy gap from different theories for the model system HLi. NAD' corresponds to the potential inverted from a density that integrates to 1.5.}
\label{tableEgsHLi}
\end{table}
For the system in which any atom carries a density distribution at its vicinity that integrates into an integer value, the step structure forms very close to the larger atom where the overlap between the densities cannot be minimised ultimately. What is evident is the fact that even the localised density integrating close to 2 is not localised in the space. Still, its difference with the orbital density happens to be localised. That's why in Fig.[\ref{rhosHLiFrom15To195}] we see that the inverted potential from all different $\rho_B(\textbf{r})$ provide a smooth curve around the H atom and in between the atoms except the potential obtained by $\int_V\rho_B(\textbf{r})d\textbf{r}=1.5$.
This latter could be a good candidate from which one can extract the most accurate energy gap.

It is essential to know that the $v^{\text{NAD}}[\rho_B,\rho_{tot}](\textbf{r})$ obtained from the $\rho_B(\textbf{r})$ in which  $\int_V\rho_B(\textbf{r})d\textbf{r}>1.5$ does not calculate very accurately the energy gap for a system in which the atoms are forming strong bounding. 
\newpage
\subsection{Homonuclear System}
	\begin{figure}[ht]
 \begin{tikzpicture}
	\node [anchor=north west] (imgA) at (-0.275\linewidth,.90\linewidth)
			{\includegraphics[width=0.8\linewidth]{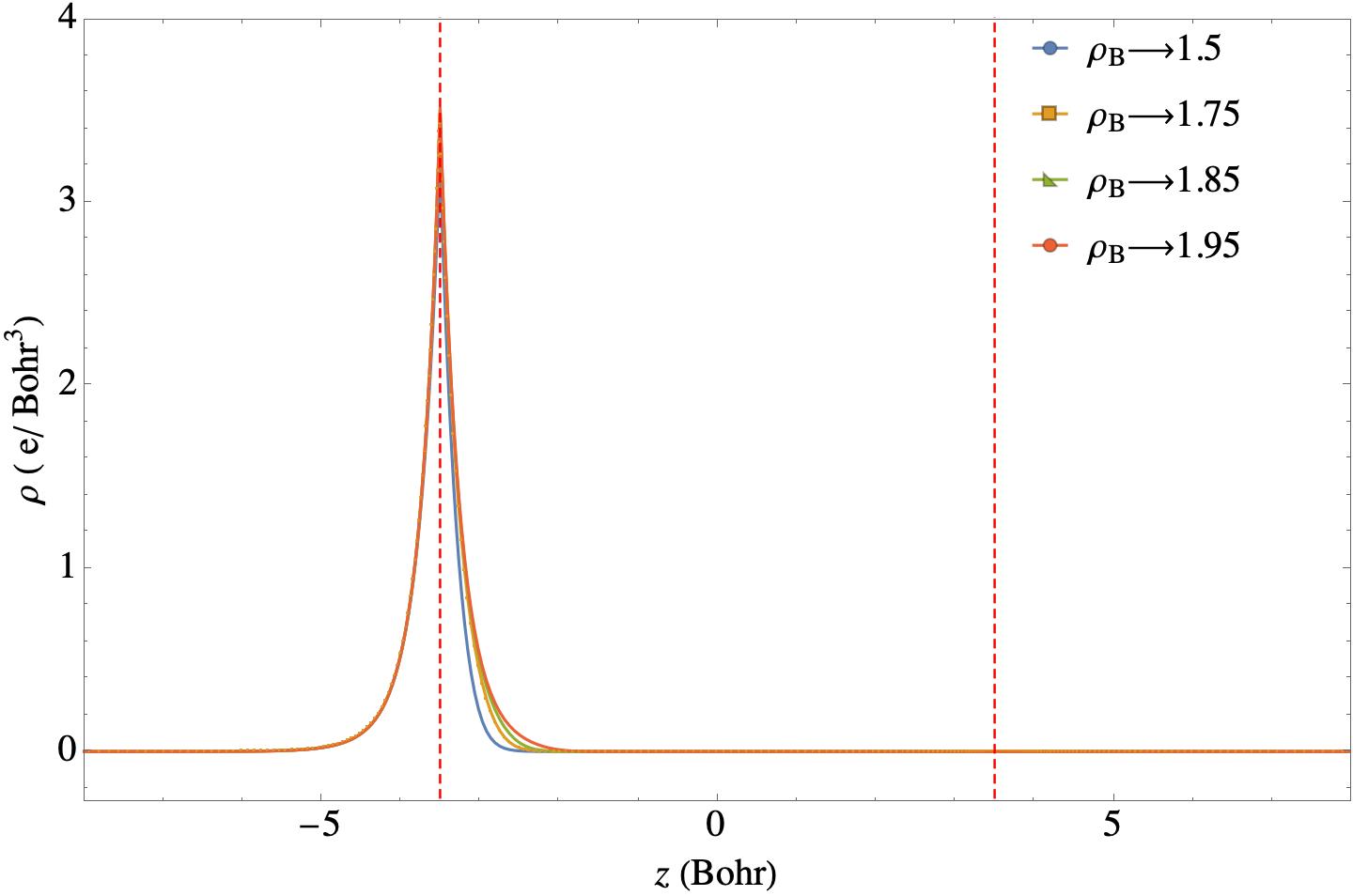}};
   \node [anchor=north west] (imgC) at (-0.275\linewidth,.35\linewidth)
            {\includegraphics[width=0.8\linewidth]{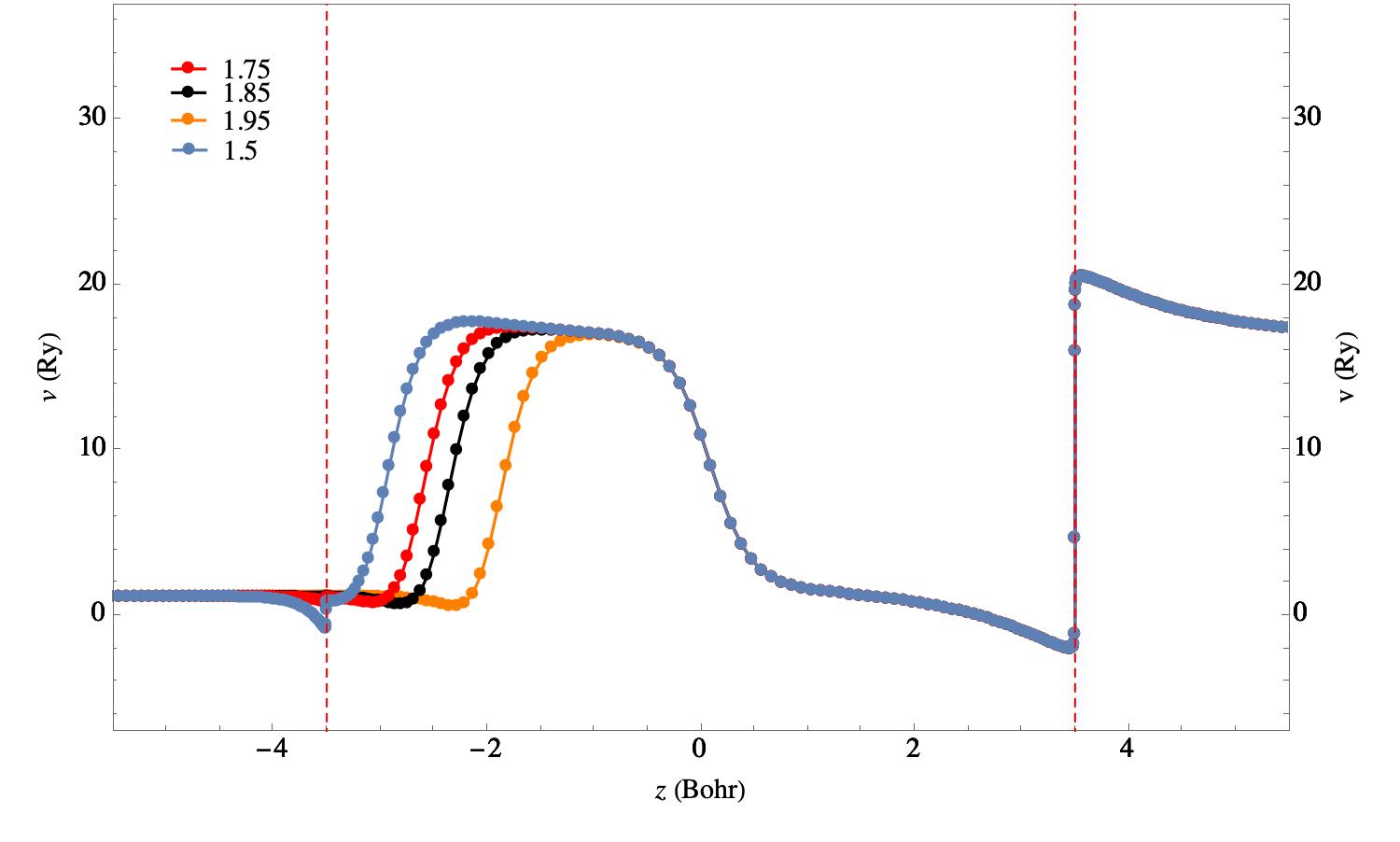}};
    \draw [anchor=north west] (-0.2\linewidth, .95\linewidth) node {(a) {\fontfamily{Arial}\selectfont {}}};
    \draw [anchor=north west] (-0.2\linewidth, .41\linewidth) node {(b) {\fontfamily{Arial}\selectfont {}}};
\end{tikzpicture}
	\captionsetup{justification=raggedright,singlelinecheck=false}
\caption{System: He-He where He is at $(0,0,-3.5)$ Bohr and other He is at $(0,0,3.5)$ Bohr; Vertical brown line: nuclei.
(a) Different partial localisation of the charge density around the He nuclei at the left side of the interatomic axis; 
(b) The exact $v^{\text{NAD}}[\rho_B,\rho_{tot}](\textbf{r})$ from analytical inversion for different partial localisation of the charge density around the He nuclei at the left side of the interatomic axis; $E_{g}\simeq 1.44$ (Ry).
}
		\label{RhosUnlocHeHe}
	\end{figure}
	
	\begin{figure}[ht]
	\centering
		\includegraphics[width=9cm]{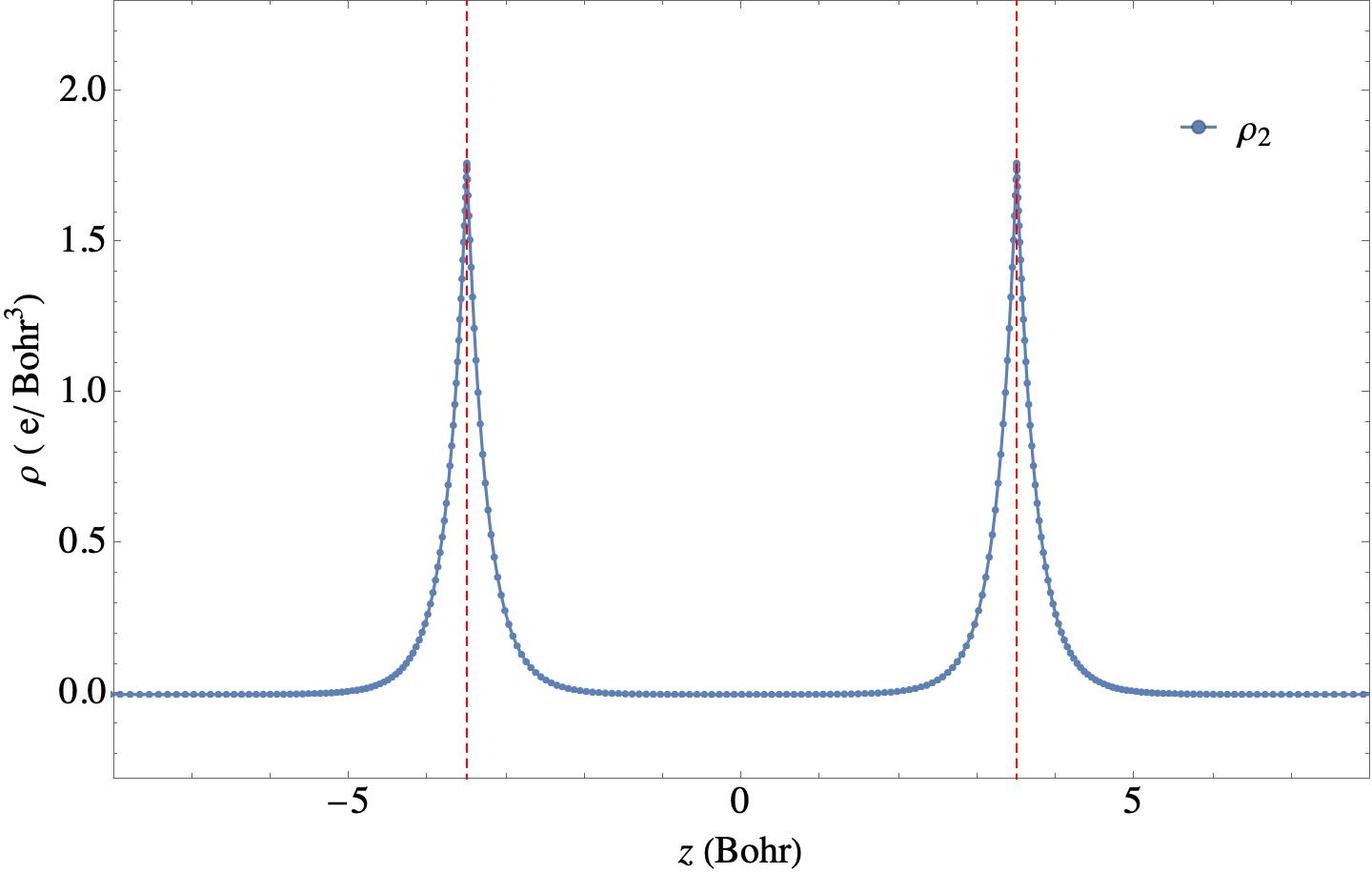}
         \captionsetup{justification=raggedright,singlelinecheck=false}
		\caption{1D representation of the $2s$ orbital $\rho_2(\textbf{r})=2|\phi_2(\textbf{r})|^2$ for the system He-He; He is sited at $(0,0,-3.5)$ Bohr and other He is located at $(0,0,3.5)$ Bohr; Vertical brown line: nuclei.}
		\label{Rho2HeHe}
	\end{figure}

	\begin{figure}[ht]
	\centering
		\includegraphics[width=9cm]{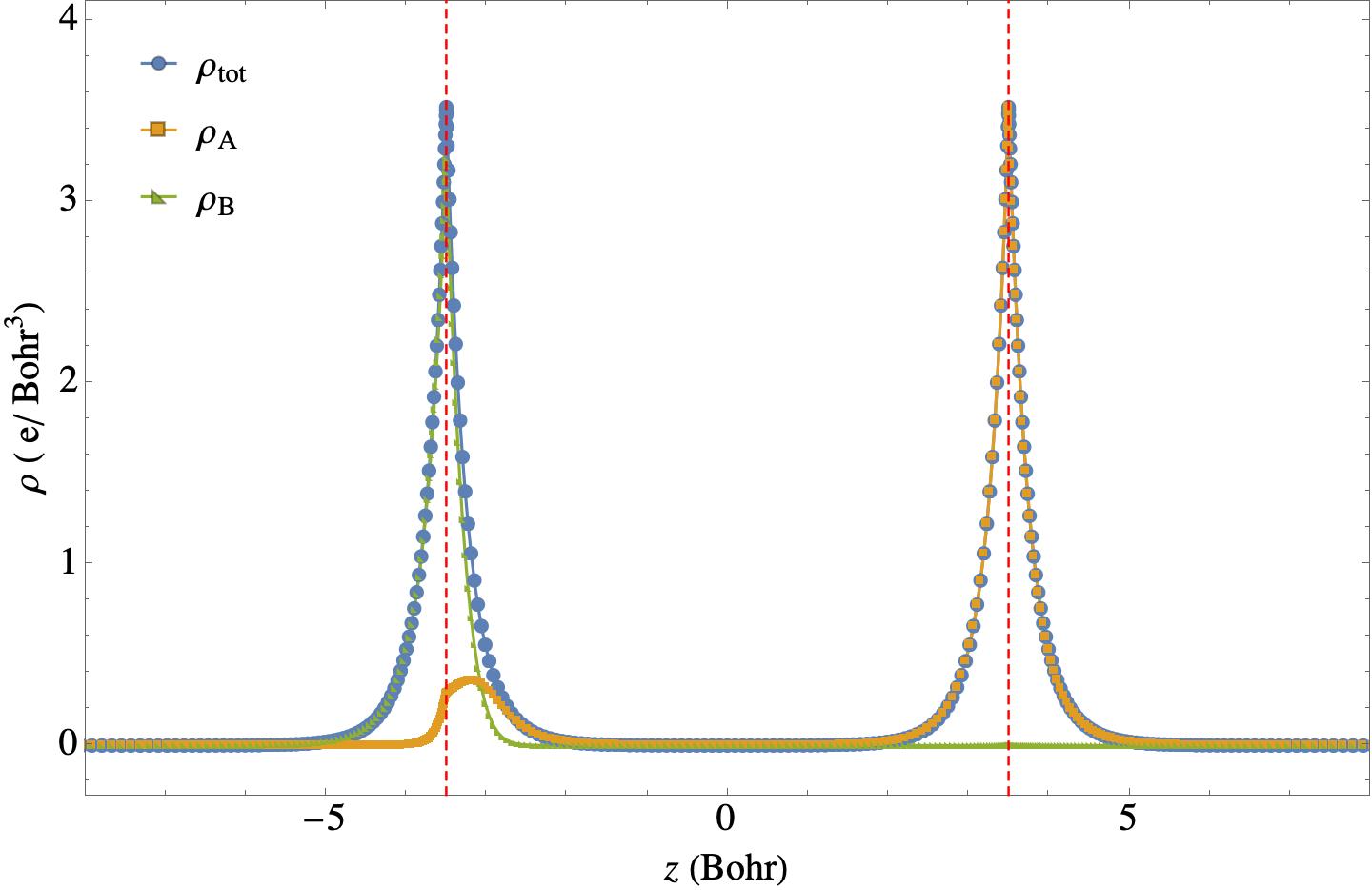}
        \captionsetup{justification=raggedright,singlelinecheck=false}
		\caption{1D representation of the $\rho_B(\textbf{r})$ , $\rho_A(\textbf{r})$and $\rho_{tot}(\textbf{r})$for the system He-He; He is sited at $(0,0,-3.5)$ Bohr and other He is located at $(0,0,3.5)$ Bohr; Vertical red line: nuclei.}
		\label{rhosHeHe15}
	\end{figure}

\begin{table}[ht]
\centering
\begin{tabular}{ |c    ||c|c|c|c| } 
\hline
Theory & \textbf{\rm HOMO} (Ry) & \textbf{\rm LUMO} (Ry) & \textbf{Gap} (Ry) \\
\hline\hline
\multirow{1}{4em}{\textbf{LDA}} & $  -1.14$ 
& $0.07$ & $1.206$ \\ 
\hline
\multirow{1}{4em}{\textbf{KLI}} & $  -1.835$ 
& $-0.280$  & $1.555$ \\ 
\hline
\multirow{1}{4em}{\textbf{NAD}} & $--$ 
&  $--$   & $1.4727$ \\
\hline
\multirow{1}{4em}{\textbf{NAD'}}   &   $--$ 
&  $--$   & \color{red}{$1.555$} \\
\hline
\end{tabular}
\captionsetup{justification=raggedright,singlelinecheck=false}
\caption{ The energy gap from different theories for the model system He-He. NAD refers to the potential inverted from the density that integrates to 2, and NAD' corresponds to the potential inverted from a density that integrates to 1.5.}
\label{tableEgsHeHe}
\end{table}

	\begin{figure}[ht]
	\centering
		\includegraphics[width=9cm]{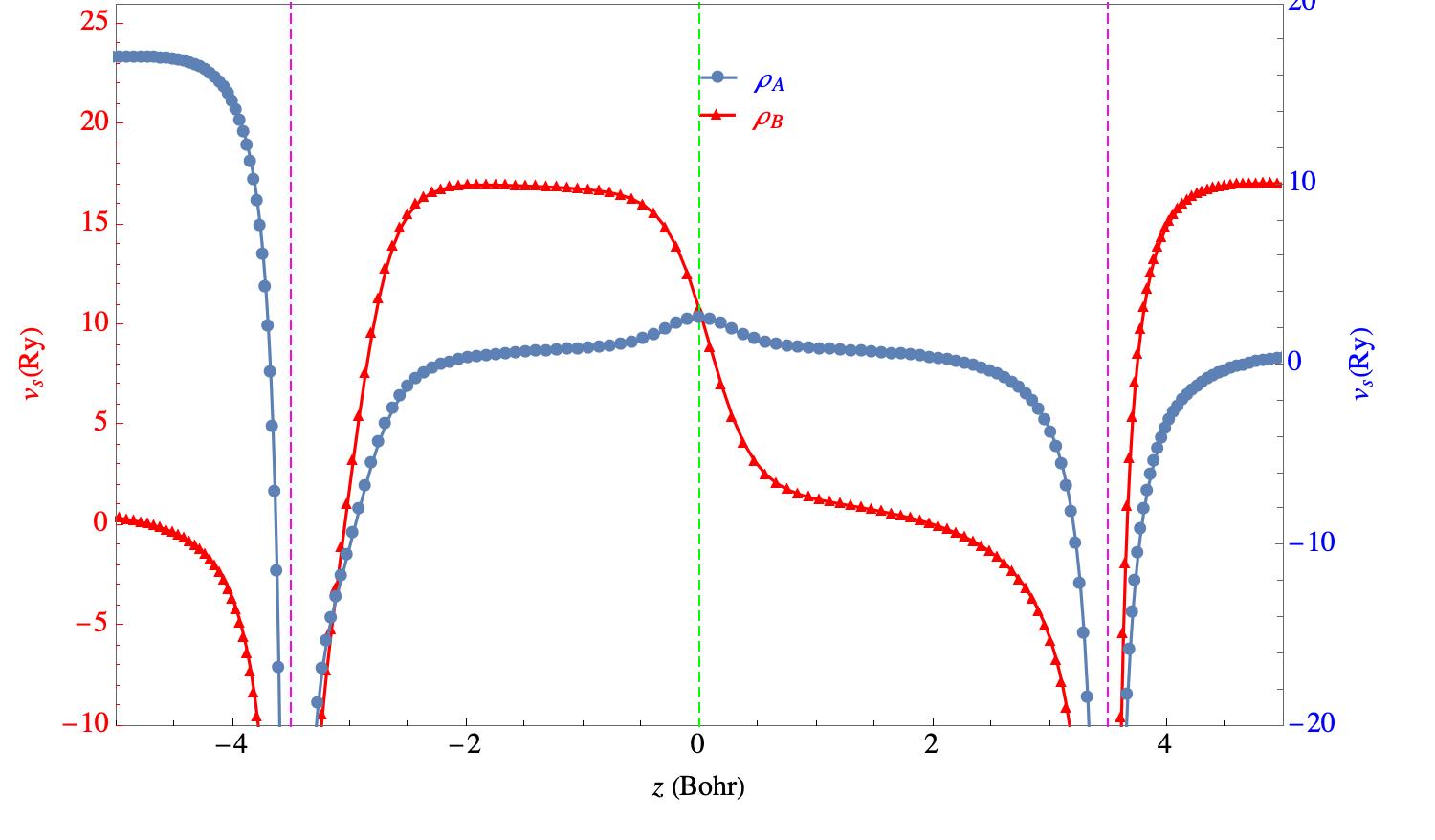}
  \captionsetup{justification=raggedright,singlelinecheck=false}
		\caption{Analytical inversion of the potential from $\rho_A(\textbf{r})$ (localised around He at the right side of the interatomic axis) and $\rho_B(\textbf{r})$ (localised around He at the left side of the interatomic axis) for partial localisation ($\int \rho_B(\textbf{r}) d\textbf{r}=1.5$) of the charge density around the He nuclei at the left side of the interatomic axis; System: He-He where one He is at $(0,0,-3.5)$ Bohr and another one is at $(0,0,3.5)$ Bohr; Vertical brown line: nuclei. The $E_{g}\simeq 1.555$ (Ry).}
		\label{HeHeVsAandB15zPlot}
	\end{figure}


\newpage
\section{Conclusion}
The exact analytically inverted non-additive potential bi-functional accurately predicts the position of the step structure. This latter is correct for both cases of integer-localisation of the charge density around nuclei and partially localised density. 

The energy gap obtained from the analytically inverted potential is strongly comparable with the one obtained from orbital related DFT model, the OEP-KLI. The difference between two energies becomes negligible when the bi-functional potential is inverted analytically from partially localised charge density. The case related to the partially localised charge density is the more realistic case that imitates the partially occupied orbitals. The partially occupied orbitals are the source of the energy gap that must be manifested on the potential curve.  This proves that the height of the SS that appeared on the curve of the potential is also accurate.

The finding in this work confirms that the low-cost exact $v^{\text{NAD}}[\rho_{B'},\rho_{tot}](\textbf{r})$ is a very reliable and accurate candidate to help the common approximations for more relevant calculations of the band gap. It has the advantage of being applicable to the other methods in order to improve them. Such application could be done both numerically within the modification of the potential with the help of the $v^{\text{NAD}}[\rho_{B'},\rho_{tot}](\textbf{r})$ in the iterative procedures or analytically modifying the $v_{\text{xc}}(\textbf{r})$. 

The $v^{\text{NAD}}[\rho_{B},\rho_{tot}](\textbf{r})$, provides the exact position of the SS, its height and the correct $E_g$.
I may suggest that this work be repeated by other ground state densities (for example the experimental GS densities or more exact theories such as full CI).
The very first guess is that the $v^{\text{NAD}}[\rho_{B},\rho_{tot}](\textbf{r})$ will provide a very accurate energy gap if one uses one of the suggested inputs.

Another interesting point to bring up is that the width of the step appearing on the $v^{\text{NAD}}[\rho_{B},\rho_{tot}](\textbf{r})$ inverted from the charge density integrating to a non-integer value must not be neglected. It is possible that some physical interpretation could be deduced from it. The first guess is the spatial probability where the system starts to attract or repulse the electron. This latter is related to the binding energy so it is essential to investigate more in this direction.

\begin{acknowledgments}
M.B. and D.S. were supported as part of the Consortium for High Energy Density Science by the U.S. Department of Energy, National Nuclear Security Administration, Minority Serving Institution Partnership Program, under Award DE-NA0003984, and by UC Merced start-up funds. Computational resources were provided by the Multi-Environment Computer for Exploration and Discovery (MERCED) cluster at UC Merced, funded by National Science Foundation Grant No. ACI-1429783.
\end{acknowledgments}


\begin{thebibliography}{58}
\expandafter\ifx\csname natexlab\endcsname\relax\def\natexlab#1{#1}\fi
\expandafter\ifx\csname bibnamefont\endcsname\relax
  \def\bibnamefont#1{#1}\fi
\expandafter\ifx\csname bibfnamefont\endcsname\relax
  \def\bibfnamefont#1{#1}\fi
\expandafter\ifx\csname citenamefont\endcsname\relax
  \def\citenamefont#1{#1}\fi
\expandafter\ifx\csname url\endcsname\relax
  \def\url#1{\texttt{#1}}\fi
\expandafter\ifx\csname urlprefix\endcsname\relax\def\urlprefix{URL }\fi
\providecommand{\bibinfo}[2]{#2}
\providecommand{\eprint}[2][]{\url{#2}}

\bibitem[{\citenamefont{Perdew et~al.}(1982)\citenamefont{Perdew, Parr, Levy,
  and Balduz~Jr}}]{perdew1982density}
\bibinfo{author}{\bibfnamefont{J.~P.} \bibnamefont{Perdew}},
  \bibinfo{author}{\bibfnamefont{R.~G.} \bibnamefont{Parr}},
  \bibinfo{author}{\bibfnamefont{M.}~\bibnamefont{Levy}}, \bibnamefont{and}
  \bibinfo{author}{\bibfnamefont{J.~L.} \bibnamefont{Balduz~Jr}},
  \bibinfo{journal}{Phys. Rev. Lett.} \textbf{\bibinfo{volume}{49}},
  \bibinfo{pages}{1691} (\bibinfo{year}{1982}).

\bibitem[{\citenamefont{Cohen et~al.}(2012)\citenamefont{Cohen,
  Mori-S\'{a}nchez, and Yang}}]{cohen2012challenges}
\bibinfo{author}{\bibfnamefont{A.~J.} \bibnamefont{Cohen}},
  \bibinfo{author}{\bibfnamefont{P.}~\bibnamefont{Mori-S\'{a}nchez}},
  \bibnamefont{and} \bibinfo{author}{\bibfnamefont{W.}~\bibnamefont{Yang}},
  \bibinfo{journal}{Chem. Rev.} \textbf{\bibinfo{volume}{112}},
  \bibinfo{pages}{289} (\bibinfo{year}{2012}).

\bibitem[{\citenamefont{Baerends et~al.}(2013)\citenamefont{Baerends,
  Gritsenko, and Van~Meer}}]{baerends2013kohn}
\bibinfo{author}{\bibfnamefont{E.~J.} \bibnamefont{Baerends}},
  \bibinfo{author}{\bibfnamefont{O.~V.} \bibnamefont{Gritsenko}},
  \bibnamefont{and} \bibinfo{author}{\bibfnamefont{R.}~\bibnamefont{Van~Meer}},
  \bibinfo{journal}{Phys. Chem. Chem. Phys.} \textbf{\bibinfo{volume}{15}},
  \bibinfo{pages}{16408} (\bibinfo{year}{2013}).

\bibitem[{\citenamefont{Mori-S{\'a}nchez and Cohen}(2014)}]{mori2014derivative}
\bibinfo{author}{\bibfnamefont{P.}~\bibnamefont{Mori-S{\'a}nchez}}
  \bibnamefont{and} \bibinfo{author}{\bibfnamefont{A.~J.} \bibnamefont{Cohen}},
  \bibinfo{journal}{Phys. Chem. Chem. Phys.} \textbf{\bibinfo{volume}{16}},
  \bibinfo{pages}{14378} (\bibinfo{year}{2014}).

\bibitem[{\citenamefont{Mosquera and Wasserman}(2014)}]{mosquera2014integer}
\bibinfo{author}{\bibfnamefont{M.~A.} \bibnamefont{Mosquera}} \bibnamefont{and}
  \bibinfo{author}{\bibfnamefont{A.}~\bibnamefont{Wasserman}},
  \bibinfo{journal}{Phys. Rev. A} \textbf{\bibinfo{volume}{89}},
  \bibinfo{pages}{052506} (\bibinfo{year}{2014}).

\bibitem[{\citenamefont{Makmal et~al.}(2011)\citenamefont{Makmal, Kuemmel, and
  Kronik}}]{makmal2011dissociation}
\bibinfo{author}{\bibfnamefont{A.}~\bibnamefont{Makmal}},
  \bibinfo{author}{\bibfnamefont{S.}~\bibnamefont{Kuemmel}}, \bibnamefont{and}
  \bibinfo{author}{\bibfnamefont{L.}~\bibnamefont{Kronik}},
  \bibinfo{journal}{Phys. Rev. A} \textbf{\bibinfo{volume}{83}},
  \bibinfo{pages}{062512} (\bibinfo{year}{2011}).

\bibitem[{\citenamefont{Van~Leeuwen et~al.}(1995)\citenamefont{Van~Leeuwen,
  Gritsenko, and Baerends}}]{van1995step}
\bibinfo{author}{\bibfnamefont{R.}~\bibnamefont{Van~Leeuwen}},
  \bibinfo{author}{\bibfnamefont{O.}~\bibnamefont{Gritsenko}},
  \bibnamefont{and} \bibinfo{author}{\bibfnamefont{E.~J.}
  \bibnamefont{Baerends}}, \bibinfo{journal}{Z. Phys. D - Atoms Molec.
  Clusters} \textbf{\bibinfo{volume}{33}}, \bibinfo{pages}{229}
  (\bibinfo{year}{1995}).

\bibitem[{\citenamefont{Helbig et~al.}(2009)\citenamefont{Helbig, Tokatly, and
  Rubio}}]{helbig2009exact}
\bibinfo{author}{\bibfnamefont{N.}~\bibnamefont{Helbig}},
  \bibinfo{author}{\bibfnamefont{I.~V.} \bibnamefont{Tokatly}},
  \bibnamefont{and} \bibinfo{author}{\bibfnamefont{A.}~\bibnamefont{Rubio}},
  \bibinfo{journal}{J. Chem. Phys.} \textbf{\bibinfo{volume}{131}},
  \bibinfo{pages}{224105} (\bibinfo{year}{2009}).

\bibitem[{\citenamefont{Gritsenko and Baerends}(1996)}]{gritsenko1996effect}
\bibinfo{author}{\bibfnamefont{O.~V.} \bibnamefont{Gritsenko}}
  \bibnamefont{and} \bibinfo{author}{\bibfnamefont{E.~J.}
  \bibnamefont{Baerends}}, \bibinfo{journal}{Phys. Rev. A}
  \textbf{\bibinfo{volume}{54}}, \bibinfo{pages}{1957} (\bibinfo{year}{1996}).

\bibitem[{\citenamefont{Maitra}(2005)}]{maitra2005undoing}
\bibinfo{author}{\bibfnamefont{N.~T.} \bibnamefont{Maitra}},
  \bibinfo{journal}{J. Chem. Phys.} \textbf{\bibinfo{volume}{122}},
  \bibinfo{pages}{234104} (\bibinfo{year}{2005}).

\bibitem[{\citenamefont{Levy and Perdew}(1985)}]{levy1985constrained}
\bibinfo{author}{\bibfnamefont{M.}~\bibnamefont{Levy}} \bibnamefont{and}
  \bibinfo{author}{\bibfnamefont{J.~P.} \bibnamefont{Perdew}}, in
  \emph{\bibinfo{booktitle}{Density functional methods in physics}}
  (\bibinfo{publisher}{Springer}, \bibinfo{year}{1985}), pp.
  \bibinfo{pages}{11--30}.

\bibitem[{\citenamefont{Almbladh and von
  Barth}(1985{\natexlab{a}})}]{almbladh1985density}
\bibinfo{author}{\bibfnamefont{C.~O.} \bibnamefont{Almbladh}} \bibnamefont{and}
  \bibinfo{author}{\bibfnamefont{U.}~\bibnamefont{von Barth}},
  \emph{\bibinfo{title}{Density Functional Methods in Physics}}, vol.
  \bibinfo{volume}{123} of \emph{\bibinfo{series}{NATO Advanced Study Institute
  Series B}} (\bibinfo{publisher}{Plenum}, \bibinfo{year}{1985}{\natexlab{a}}),
  \bibinfo{note}{eds. R. M. Dreizler and J. da Providencia}.

\bibitem[{\citenamefont{Gritsenko and Baerends}(2006)}]{gritsenko2006correct}
\bibinfo{author}{\bibfnamefont{O.}~\bibnamefont{Gritsenko}} \bibnamefont{and}
  \bibinfo{author}{\bibfnamefont{E.~J.} \bibnamefont{Baerends}},
  \bibinfo{journal}{Int. J. Quantum Chem.} \textbf{\bibinfo{volume}{106}},
  \bibinfo{pages}{3167} (\bibinfo{year}{2006}).

\bibitem[{\citenamefont{Ruzsinszky et~al.}(2006)\citenamefont{Ruzsinszky,
  Perdew, Csonka, Vydrov, and Scuseria}}]{ruzsinszky2006spurious}
\bibinfo{author}{\bibfnamefont{A.}~\bibnamefont{Ruzsinszky}},
  \bibinfo{author}{\bibfnamefont{J.~P.} \bibnamefont{Perdew}},
  \bibinfo{author}{\bibfnamefont{G.~I.} \bibnamefont{Csonka}},
  \bibinfo{author}{\bibfnamefont{O.~A.} \bibnamefont{Vydrov}},
  \bibnamefont{and} \bibinfo{author}{\bibfnamefont{G.~E.}
  \bibnamefont{Scuseria}}, \bibinfo{journal}{J. Chem. Phys.}
  \textbf{\bibinfo{volume}{125}}, \bibinfo{pages}{194112}
  (\bibinfo{year}{2006}).

\bibitem[{\citenamefont{Gritsenko et~al.}(1996)\citenamefont{Gritsenko,
  Van~Leeuwen, and Baerends}}]{gritsenko1996molecular}
\bibinfo{author}{\bibfnamefont{O.~V.} \bibnamefont{Gritsenko}},
  \bibinfo{author}{\bibfnamefont{R.}~\bibnamefont{Van~Leeuwen}},
  \bibnamefont{and} \bibinfo{author}{\bibfnamefont{E.~J.}
  \bibnamefont{Baerends}}, \bibinfo{journal}{J. Chem. Phys.}
  \textbf{\bibinfo{volume}{104}}, \bibinfo{pages}{8535} (\bibinfo{year}{1996}).

\bibitem[{\citenamefont{Karolewski et~al.}(2009)\citenamefont{Karolewski,
  Armiento, and Kummel}}]{karolewski2009polarizabilities}
\bibinfo{author}{\bibfnamefont{A.}~\bibnamefont{Karolewski}},
  \bibinfo{author}{\bibfnamefont{R.}~\bibnamefont{Armiento}}, \bibnamefont{and}
  \bibinfo{author}{\bibfnamefont{S.}~\bibnamefont{Kummel}},
  \bibinfo{journal}{J. Chem. Theory Comput.} \textbf{\bibinfo{volume}{5}},
  \bibinfo{pages}{712} (\bibinfo{year}{2009}).

\bibitem[{\citenamefont{Tempel et~al.}(2009)\citenamefont{Tempel, Martinez, and
  Maitra}}]{tempel2009revisiting}
\bibinfo{author}{\bibfnamefont{D.~G.} \bibnamefont{Tempel}},
  \bibinfo{author}{\bibfnamefont{T.~J.} \bibnamefont{Martinez}},
  \bibnamefont{and} \bibinfo{author}{\bibfnamefont{N.~T.}
  \bibnamefont{Maitra}}, \bibinfo{journal}{J. Chem. Theory Comput.}
  \textbf{\bibinfo{volume}{5}}, \bibinfo{pages}{770} (\bibinfo{year}{2009}).

\bibitem[{\citenamefont{Hellgren et~al.}(2012)\citenamefont{Hellgren, Rohr, and
  Gross}}]{hellgren2012correlation}
\bibinfo{author}{\bibfnamefont{M.}~\bibnamefont{Hellgren}},
  \bibinfo{author}{\bibfnamefont{D.~R.} \bibnamefont{Rohr}}, \bibnamefont{and}
  \bibinfo{author}{\bibfnamefont{E.~K.~U.} \bibnamefont{Gross}},
  \bibinfo{journal}{J. Chem. Phys.} \textbf{\bibinfo{volume}{136}},
  \bibinfo{pages}{034106} (\bibinfo{year}{2012}).

\bibitem[{\citenamefont{Kraisler and Kronik}(2015)}]{kraisler2015elimination}
\bibinfo{author}{\bibfnamefont{E.}~\bibnamefont{Kraisler}} \bibnamefont{and}
  \bibinfo{author}{\bibfnamefont{L.}~\bibnamefont{Kronik}},
  \bibinfo{journal}{Phys. Rev. A} \textbf{\bibinfo{volume}{91}},
  \bibinfo{pages}{032504} (\bibinfo{year}{2015}).

\bibitem[{\citenamefont{Buijse}(1989)}]{buijse1989j}
\bibinfo{author}{\bibfnamefont{M.~A.} \bibnamefont{Buijse}},
  \bibinfo{journal}{Phys. Rev. A} \textbf{\bibinfo{volume}{40}},
  \bibinfo{pages}{4190} (\bibinfo{year}{1989}).

\bibitem[{\citenamefont{Hodgson et~al.}(2016)\citenamefont{Hodgson, Ramsden,
  and Godby}}]{hodgson2016origin}
\bibinfo{author}{\bibfnamefont{M.~J.~P.} \bibnamefont{Hodgson}},
  \bibinfo{author}{\bibfnamefont{J.~D.} \bibnamefont{Ramsden}},
  \bibnamefont{and} \bibinfo{author}{\bibfnamefont{R.~W.} \bibnamefont{Godby}},
  \bibinfo{journal}{Phys. Rev. B} \textbf{\bibinfo{volume}{93}},
  \bibinfo{pages}{155146} (\bibinfo{year}{2016}).

\bibitem[{\citenamefont{Hodgson et~al.}(2017)\citenamefont{Hodgson, Kraisler,
  Schild, and Gross}}]{hodgson2017interatomic}
\bibinfo{author}{\bibfnamefont{M.~J.~P.} \bibnamefont{Hodgson}},
  \bibinfo{author}{\bibfnamefont{E.}~\bibnamefont{Kraisler}},
  \bibinfo{author}{\bibfnamefont{A.}~\bibnamefont{Schild}}, \bibnamefont{and}
  \bibinfo{author}{\bibfnamefont{E.~K.~U.} \bibnamefont{Gross}},
  \bibinfo{journal}{J. Phys. Chem. Lett.} \textbf{\bibinfo{volume}{8}},
  \bibinfo{pages}{5974} (\bibinfo{year}{2017}).

\bibitem[{\citenamefont{Perdew and Wang}(1992)}]{perdew1992accurate}
\bibinfo{author}{\bibfnamefont{J.~P.} \bibnamefont{Perdew}} \bibnamefont{and}
  \bibinfo{author}{\bibfnamefont{Y.}~\bibnamefont{Wang}},
  \bibinfo{journal}{Phys. Rev. B} \textbf{\bibinfo{volume}{45}},
  \bibinfo{pages}{13244} (\bibinfo{year}{1992}).

\bibitem[{\citenamefont{Becke}(1988)}]{becke1988density}
\bibinfo{author}{\bibfnamefont{A.~D.} \bibnamefont{Becke}},
  \bibinfo{journal}{Phys. Rev. A} \textbf{\bibinfo{volume}{38}},
  \bibinfo{pages}{3098} (\bibinfo{year}{1988}).

\bibitem[{\citenamefont{Lee et~al.}(1988)\citenamefont{Lee, Yang, and
  Parr}}]{lee1988development}
\bibinfo{author}{\bibfnamefont{C.}~\bibnamefont{Lee}},
  \bibinfo{author}{\bibfnamefont{W.}~\bibnamefont{Yang}}, \bibnamefont{and}
  \bibinfo{author}{\bibfnamefont{R.~G.} \bibnamefont{Parr}},
  \bibinfo{journal}{Phys. Rev. B} \textbf{\bibinfo{volume}{37}},
  \bibinfo{pages}{785} (\bibinfo{year}{1988}).

\bibitem[{\citenamefont{Perdew et~al.}(1996)\citenamefont{Perdew, Burke, and
  Ernzerhof}}]{perdew1996generalized}
\bibinfo{author}{\bibfnamefont{J.~P.} \bibnamefont{Perdew}},
  \bibinfo{author}{\bibfnamefont{K.}~\bibnamefont{Burke}}, \bibnamefont{and}
  \bibinfo{author}{\bibfnamefont{M.}~\bibnamefont{Ernzerhof}},
  \bibinfo{journal}{Phys. Rev. Lett.} \textbf{\bibinfo{volume}{77}},
  \bibinfo{pages}{3865} (\bibinfo{year}{1996}).

\bibitem[{\citenamefont{Yang et~al.}(2012)\citenamefont{Yang, Cohen, and
  Mori-Sanchez}}]{yang2012derivative}
\bibinfo{author}{\bibfnamefont{W.}~\bibnamefont{Yang}},
  \bibinfo{author}{\bibfnamefont{A.~J.} \bibnamefont{Cohen}}, \bibnamefont{and}
  \bibinfo{author}{\bibfnamefont{P.}~\bibnamefont{Mori-Sanchez}},
  \bibinfo{journal}{J. Chem. Phys.} \textbf{\bibinfo{volume}{136}},
  \bibinfo{pages}{204111} (\bibinfo{year}{2012}).

\bibitem[{\citenamefont{Levy et~al.}(1984)\citenamefont{Levy, Perdew, and
  Sahni}}]{levy1984exact}
\bibinfo{author}{\bibfnamefont{M.}~\bibnamefont{Levy}},
  \bibinfo{author}{\bibfnamefont{J.~P.} \bibnamefont{Perdew}},
  \bibnamefont{and} \bibinfo{author}{\bibfnamefont{V.}~\bibnamefont{Sahni}},
  \bibinfo{journal}{Phys. Rev. A} \textbf{\bibinfo{volume}{30}},
  \bibinfo{pages}{2745} (\bibinfo{year}{1984}).

\bibitem[{\citenamefont{Perdew and Levy}(1997)}]{perdew1997comment}
\bibinfo{author}{\bibfnamefont{J.~P.} \bibnamefont{Perdew}} \bibnamefont{and}
  \bibinfo{author}{\bibfnamefont{M.}~\bibnamefont{Levy}},
  \bibinfo{journal}{Phys. Rev. B} \textbf{\bibinfo{volume}{56}},
  \bibinfo{pages}{16021} (\bibinfo{year}{1997}).

\bibitem[{\citenamefont{Harbola}(1999)}]{harbola1999relationship}
\bibinfo{author}{\bibfnamefont{M.~K.} \bibnamefont{Harbola}},
  \bibinfo{journal}{Phys. Rev. B} \textbf{\bibinfo{volume}{60}},
  \bibinfo{pages}{4545} (\bibinfo{year}{1999}).

\bibitem[{\citenamefont{Perdew}(1985)}]{perdew1985density}
\bibinfo{author}{\bibfnamefont{J.~P.} \bibnamefont{Perdew}},
  \emph{\bibinfo{title}{Density Functional Methods in Physics}}, vol.
  \bibinfo{volume}{123} of \emph{\bibinfo{series}{NATO Advanced Study Institute
  Series B}} (\bibinfo{publisher}{Plenum}, \bibinfo{year}{1985}),
  \bibinfo{note}{eds. R. M. Dreizler and J. da Providencia}.

\bibitem[{\citenamefont{Hellgren and
  Gross}(2012)}]{hellgren2012discontinuities}
\bibinfo{author}{\bibfnamefont{M.}~\bibnamefont{Hellgren}} \bibnamefont{and}
  \bibinfo{author}{\bibfnamefont{E.~K.~U.} \bibnamefont{Gross}},
  \bibinfo{journal}{Phys. Rev. A} \textbf{\bibinfo{volume}{85}},
  \bibinfo{pages}{022514} (\bibinfo{year}{2012}).

\bibitem[{\citenamefont{Ben{\'\i}tez and Proetto}(2016)}]{benitez2016kohn}
\bibinfo{author}{\bibfnamefont{A.}~\bibnamefont{Ben{\'\i}tez}}
  \bibnamefont{and} \bibinfo{author}{\bibfnamefont{C.~R.}
  \bibnamefont{Proetto}}, \bibinfo{journal}{Phys. Rev. A}
  \textbf{\bibinfo{volume}{94}}, \bibinfo{pages}{052506}
  (\bibinfo{year}{2016}).

\bibitem[{\citenamefont{Hofmann and K{\"u}mmel}(2012)}]{hofmann2012integer}
\bibinfo{author}{\bibfnamefont{D.}~\bibnamefont{Hofmann}} \bibnamefont{and}
  \bibinfo{author}{\bibfnamefont{S.}~\bibnamefont{K{\"u}mmel}},
  \bibinfo{journal}{Phys. Rev. B} \textbf{\bibinfo{volume}{86}},
  \bibinfo{pages}{201109} (\bibinfo{year}{2012}).

\bibitem[{\citenamefont{Perdew}(1990)}]{perdew1990size}
\bibinfo{author}{\bibfnamefont{J.~P.} \bibnamefont{Perdew}}, in
  \emph{\bibinfo{booktitle}{Advances in Quantum Chemistry}}
  (\bibinfo{publisher}{Elsevier}, \bibinfo{year}{1990}),
  vol.~\bibinfo{volume}{21}, pp. \bibinfo{pages}{113--134}.

\bibitem[{\citenamefont{Sagvolden and
  Perdew}(2008)}]{sagvolden2008discontinuity}
\bibinfo{author}{\bibfnamefont{E.}~\bibnamefont{Sagvolden}} \bibnamefont{and}
  \bibinfo{author}{\bibfnamefont{J.~P.} \bibnamefont{Perdew}},
  \bibinfo{journal}{Phys. Rev. A} \textbf{\bibinfo{volume}{77}},
  \bibinfo{pages}{012517} (\bibinfo{year}{2008}).

\bibitem[{\citenamefont{Fuks et~al.}(2011)\citenamefont{Fuks, Rubio, and
  Maitra}}]{fuks2011charge}
\bibinfo{author}{\bibfnamefont{J.~I.} \bibnamefont{Fuks}},
  \bibinfo{author}{\bibfnamefont{A.}~\bibnamefont{Rubio}}, \bibnamefont{and}
  \bibinfo{author}{\bibfnamefont{N.~T.} \bibnamefont{Maitra}},
  \bibinfo{journal}{Phys. Rev. A} \textbf{\bibinfo{volume}{83}},
  \bibinfo{pages}{042501} (\bibinfo{year}{2011}).

\bibitem[{\citenamefont{Hellgren and Gould}(2019)}]{gould2014delocalization}
\bibinfo{author}{\bibfnamefont{M.}~\bibnamefont{Hellgren}} \bibnamefont{and}
  \bibinfo{author}{\bibfnamefont{T.}~\bibnamefont{Gould}}, \bibinfo{journal}{J.
  Chem. Theory Comput.} \textbf{\bibinfo{volume}{15}}, \bibinfo{pages}{4907}
  (\bibinfo{year}{2019}).

\bibitem[{\citenamefont{Nafziger and Wasserman}(2015)}]{nafziger2015fragment}
\bibinfo{author}{\bibfnamefont{J.}~\bibnamefont{Nafziger}} \bibnamefont{and}
  \bibinfo{author}{\bibfnamefont{A.}~\bibnamefont{Wasserman}},
  \bibinfo{journal}{J. Chem. Phys.} \textbf{\bibinfo{volume}{143}},
  \bibinfo{pages}{234105} (\bibinfo{year}{2015}).

\bibitem[{\citenamefont{Kohut et~al.}(2016)\citenamefont{Kohut, Polgar, and
  Staroverov}}]{kohut2016origin}
\bibinfo{author}{\bibfnamefont{S.~V.} \bibnamefont{Kohut}},
  \bibinfo{author}{\bibfnamefont{A.~M.} \bibnamefont{Polgar}},
  \bibnamefont{and} \bibinfo{author}{\bibfnamefont{V.~N.}
  \bibnamefont{Staroverov}}, \bibinfo{journal}{Phys. Chem. Chem. Phys.}
  \textbf{\bibinfo{volume}{18}}, \bibinfo{pages}{20938} (\bibinfo{year}{2016}).

\bibitem[{\citenamefont{Komsa and Staroverov}(2016)}]{komsa2016elimination}
\bibinfo{author}{\bibfnamefont{D.~N.} \bibnamefont{Komsa}} \bibnamefont{and}
  \bibinfo{author}{\bibfnamefont{V.~N.} \bibnamefont{Staroverov}},
  \bibinfo{journal}{J. Chem. Theory Comput.} \textbf{\bibinfo{volume}{12}},
  \bibinfo{pages}{5361} (\bibinfo{year}{2016}).

\bibitem[{\citenamefont{Li et~al.}(2015)\citenamefont{Li, Zheng, Cohen,
  Mori-S{\'a}nchez, and Yang}}]{li2015local}
\bibinfo{author}{\bibfnamefont{C.}~\bibnamefont{Li}},
  \bibinfo{author}{\bibfnamefont{X.}~\bibnamefont{Zheng}},
  \bibinfo{author}{\bibfnamefont{A.~J.} \bibnamefont{Cohen}},
  \bibinfo{author}{\bibfnamefont{P.}~\bibnamefont{Mori-S{\'a}nchez}},
  \bibnamefont{and} \bibinfo{author}{\bibfnamefont{W.}~\bibnamefont{Yang}},
  \bibinfo{journal}{Phys. Rev. Lett.} \textbf{\bibinfo{volume}{114}},
  \bibinfo{pages}{053001} (\bibinfo{year}{2015}).

\bibitem[{\citenamefont{K{\"u}mmel and Kronik}(2008)}]{kummel2008orbital}
\bibinfo{author}{\bibfnamefont{S.}~\bibnamefont{K{\"u}mmel}} \bibnamefont{and}
  \bibinfo{author}{\bibfnamefont{L.}~\bibnamefont{Kronik}},
  \bibinfo{journal}{Rev. Mod. Phys.} \textbf{\bibinfo{volume}{80}},
  \bibinfo{pages}{3} (\bibinfo{year}{2008}).

\bibitem[{\citenamefont{Sharp and Horton}(1953)}]{sharp1953variational}
\bibinfo{author}{\bibfnamefont{R.~T.} \bibnamefont{Sharp}} \bibnamefont{and}
  \bibinfo{author}{\bibfnamefont{G.~K.} \bibnamefont{Horton}},
  \bibinfo{journal}{Phys. Rev.} \textbf{\bibinfo{volume}{90}},
  \bibinfo{pages}{317} (\bibinfo{year}{1953}).

\bibitem[{\citenamefont{Talman and Shadwick}(1976)}]{talman1976optimized}
\bibinfo{author}{\bibfnamefont{J.~D.} \bibnamefont{Talman}} \bibnamefont{and}
  \bibinfo{author}{\bibfnamefont{W.~F.} \bibnamefont{Shadwick}},
  \bibinfo{journal}{Phys. Rev. A} \textbf{\bibinfo{volume}{14}},
  \bibinfo{pages}{36} (\bibinfo{year}{1976}).

\bibitem[{\citenamefont{Grabo et~al.}(1997)\citenamefont{Grabo, Kreibich, and
  Gross}}]{grabo1997optimized}
\bibinfo{author}{\bibfnamefont{T.}~\bibnamefont{Grabo}},
  \bibinfo{author}{\bibfnamefont{T.}~\bibnamefont{Kreibich}}, \bibnamefont{and}
  \bibinfo{author}{\bibfnamefont{E.~K.~U.} \bibnamefont{Gross}},
  \bibinfo{journal}{Mol. Eng.} \textbf{\bibinfo{volume}{7}},
  \bibinfo{pages}{27} (\bibinfo{year}{1997}).

\bibitem[{\citenamefont{Engel}(2003)}]{engel2003orbital}
\bibinfo{author}{\bibfnamefont{E.}~\bibnamefont{Engel}}, \bibinfo{journal}{A
  primer in density functional theory} pp. \bibinfo{pages}{56--122}
  (\bibinfo{year}{2003}).

\bibitem[{\citenamefont{Krieger et~al.}(1992)\citenamefont{Krieger, Li, and
  Iafrate}}]{krieger1992construction}
\bibinfo{author}{\bibfnamefont{J.~B.} \bibnamefont{Krieger}},
  \bibinfo{author}{\bibfnamefont{Y.}~\bibnamefont{Li}}, \bibnamefont{and}
  \bibinfo{author}{\bibfnamefont{G.~J.} \bibnamefont{Iafrate}},
  \bibinfo{journal}{Phys. Rev. A} \textbf{\bibinfo{volume}{45}},
  \bibinfo{pages}{101} (\bibinfo{year}{1992}).

\bibitem[{\citenamefont{Almbladh and von
  Barth}(1985{\natexlab{b}})}]{almbladh1985exact}
\bibinfo{author}{\bibfnamefont{C.-O.} \bibnamefont{Almbladh}} \bibnamefont{and}
  \bibinfo{author}{\bibfnamefont{U.}~\bibnamefont{von Barth}},
  \bibinfo{journal}{Phys. Rev. B} \textbf{\bibinfo{volume}{31}},
  \bibinfo{pages}{3231} (\bibinfo{year}{1985}{\natexlab{b}}).

\bibitem[{\citenamefont{Della~Sala and
  G{\"o}rling}(2002)}]{della2002asymptotic}
\bibinfo{author}{\bibfnamefont{F.}~\bibnamefont{Della~Sala}} \bibnamefont{and}
  \bibinfo{author}{\bibfnamefont{A.}~\bibnamefont{G{\"o}rling}},
  \bibinfo{journal}{Phys. Rev. Lett.} \textbf{\bibinfo{volume}{89}},
  \bibinfo{pages}{033003} (\bibinfo{year}{2002}).

\bibitem[{\citenamefont{Perdew and Zunger}(1981)}]{perdew1981self}
\bibinfo{author}{\bibfnamefont{J.~P.} \bibnamefont{Perdew}} \bibnamefont{and}
  \bibinfo{author}{\bibfnamefont{A.}~\bibnamefont{Zunger}},
  \bibinfo{journal}{Phys. Rev. B} \textbf{\bibinfo{volume}{23}},
  \bibinfo{pages}{5048} (\bibinfo{year}{1981}).

\bibitem[{\citenamefont{Vydrov et~al.}(2006)\citenamefont{Vydrov, Scuseria,
  Perdew, Ruzsinszky, and Csonka}}]{vydrov2006scaling}
\bibinfo{author}{\bibfnamefont{O.~A.} \bibnamefont{Vydrov}},
  \bibinfo{author}{\bibfnamefont{G.~E.} \bibnamefont{Scuseria}},
  \bibinfo{author}{\bibfnamefont{J.~P.} \bibnamefont{Perdew}},
  \bibinfo{author}{\bibfnamefont{A.}~\bibnamefont{Ruzsinszky}},
  \bibnamefont{and} \bibinfo{author}{\bibfnamefont{G.~I.}
  \bibnamefont{Csonka}}, \bibinfo{journal}{J. Chem. Phys.}
  \textbf{\bibinfo{volume}{124}}, \bibinfo{pages}{094108}
  (\bibinfo{year}{2006}).

\bibitem[{\citenamefont{Banafsheh et~al.}(2022)\citenamefont{Banafsheh,
  Wesolowski, Gould, Kronik, and Strubbe}}]{banafsheh2022nuclear}
\bibinfo{author}{\bibfnamefont{M.}~\bibnamefont{Banafsheh}},
  \bibinfo{author}{\bibfnamefont{T.~A.} \bibnamefont{Wesolowski}},
  \bibinfo{author}{\bibfnamefont{T.}~\bibnamefont{Gould}},
  \bibinfo{author}{\bibfnamefont{L.}~\bibnamefont{Kronik}}, \bibnamefont{and}
  \bibinfo{author}{\bibfnamefont{D.~A.} \bibnamefont{Strubbe}},
  \bibinfo{journal}{Phys. Rev. A} \textbf{\bibinfo{volume}{106}},
  \bibinfo{pages}{042812} (\bibinfo{year}{2022}).

\bibitem[{\citenamefont{Banafsheh and A.~Wesolowski}(2018)}]{Banafsheh}
\bibinfo{author}{\bibfnamefont{M.}~\bibnamefont{Banafsheh}} \bibnamefont{and}
  \bibinfo{author}{\bibfnamefont{T.}~\bibnamefont{A.~Wesolowski}},
  \bibinfo{journal}{Int. J. Quantum Chem.} \textbf{\bibinfo{volume}{118}},
  \bibinfo{pages}{e25410} (\bibinfo{year}{2018}).

\bibitem[{\citenamefont{Makmal et~al.}(2009)\citenamefont{Makmal, K\"{u}mmel,
  and Kronik}}]{makmal2009fully}
\bibinfo{author}{\bibfnamefont{A.}~\bibnamefont{Makmal}},
  \bibinfo{author}{\bibfnamefont{S.}~\bibnamefont{K\"{u}mmel}},
  \bibnamefont{and} \bibinfo{author}{\bibfnamefont{L.}~\bibnamefont{Kronik}},
  \bibinfo{journal}{J. Chem. Theory Comput.} \textbf{\bibinfo{volume}{5}},
  \bibinfo{pages}{1731} (\bibinfo{year}{2009}).

\bibitem[{\citenamefont{Fornberg}(1988)}]{fornberg1988generation}
\bibinfo{author}{\bibfnamefont{B.}~\bibnamefont{Fornberg}},
  \bibinfo{journal}{Math. Comput.} \textbf{\bibinfo{volume}{51}},
  \bibinfo{pages}{699} (\bibinfo{year}{1988}).

\bibitem[{\citenamefont{Beck}(2000)}]{beck2000real}
\bibinfo{author}{\bibfnamefont{T.~L.} \bibnamefont{Beck}},
  \bibinfo{journal}{Rev. Mod. Phys.} \textbf{\bibinfo{volume}{72}},
  \bibinfo{pages}{1041} (\bibinfo{year}{2000}).

\bibitem[{\citenamefont{Ceperley and Alder}(1980)}]{ceperley1980ground}
\bibinfo{author}{\bibfnamefont{D.~M.} \bibnamefont{Ceperley}} \bibnamefont{and}
  \bibinfo{author}{\bibfnamefont{B.~J.} \bibnamefont{Alder}},
  \bibinfo{journal}{Phys. Rev. Lett.} \textbf{\bibinfo{volume}{45}},
  \bibinfo{pages}{566} (\bibinfo{year}{1980}).

\end{thebibliography}


\end{document}